\documentclass[10pt,a4paper,twocolumn]{article}
\usepackage[a4paper,margin=1in,columnsep=25pt]{geometry}
\usepackage[utf8]{inputenc}
\usepackage{amssymb,amsfonts,amsmath}
\usepackage{graphicx}
\usepackage{booktabs,tabularx,multirow}
\usepackage{float}
\usepackage[table]{xcolor}
\usepackage{caption,subcaption}
\usepackage{enumitem}
\usepackage{csquotes}
\usepackage{tikz}
\usetikzlibrary{positioning,calc}
\usepackage[most]{tcolorbox}
\usepackage{authblk}
\usepackage{hyperref}
\usepackage{fancyhdr}
\begin{document}

\definecolor{purple}{HTML}{6d4aff}
\newtcolorbox{disclaimerbox}{
    colback=purple!5,
    colframe=purple,
    boxrule=0.8pt,
    arc=3pt,
    left=10pt, right=10pt, top=4pt, bottom=4pt,
    fontupper=\scriptsize
}

\newcommand{\rebuttal}[1]{\textcolor{blue}{#1}}

\newcommand{\DAMI}{DM} 
\newcommand{\PDS}{PDS} 
\newcommand{\todo}[1]{\textcolor{myred}{\textit{todo}: \textit{#1}}}

\long\def\acks#1{\vskip 0.3in\noindent{\large\bf Acknowledgements}\vskip 0.2in
	\noindent #1}

\long\def\ethics#1{\vskip 0.3in\noindent{\large\bf Ethical Standards}\vskip 0.2in
\noindent #1}

\long\def\coi#1{\vskip 0.3in\noindent{\large\bf Conflicts of Interest}\vskip 0.2in
	\noindent #1}

\long\def\data#1{\vskip 0.3in\noindent{\large\bf Data availability}\vskip 0.2in
  \noindent #1}

\title{Data annotations as pedagogical hints: \\ \large from subjective labels to critical thinking}

\author[1,2]{Ralf Raumanns}
\author[3]{Theresa Elstner}
\author[1]{Louis Ferger-Andrews}
\author[4]{Louise M. Carlsen}
\author[3,4]{Martin Potthast}
\author[1]{Gerard Schouten}
\author[2]{Josien P. W. Pluim}
\author[5]{Veronika Cheplygina}

\affil[1]{Fontys University of Applied Sciences, Venlo, The Netherlands}
\affil[2]{Eindhoven University of Technology, Eindhoven, The Netherlands}
\affil[3]{University of Kassel and hessian.AI, Kassel, Germany}
\affil[4]{ScaDS.AI, Dreden/Leipzig, Germany}
\affil[5]{IT University of Copenhagen, Denmark}

\date{}

\twocolumn[
  \maketitle
  \begin{@twocolumnfalse}
\begin{abstract}
Machine learning courses often rely on pre-labelled datasets. Yet the subjectivity inherent in human annotation remains invisible to students. At worst, this teaching approach produces learners with an overly trusting and potentially harmful perspective. They develop a false sense of objectivity in data and AI models. This comes at the expense of valuing interpretive diversity and contestability of algorithmic outputs. In this paper, we show that assigning learners a manual data annotation task leads to an understanding of the subjective nature of human-given labels.

More precisely, we explore the extent to which manual annotation of images teaches students how human judgement shapes AI training. We investigate (1) How does group annotation affect perceptions of data quality, bias, and fairness? (2) What are the strengths and limitations of this teaching concept?

We implemented an annotation activity at two universities: Fontys (NL, bachelor's) and IT University Copenhagen (DK, bachelor's). Students annotated skin lesion images for hair coverage on a 3-point scale. We then collected surveys from 43 students. These measure students’ understanding of annotation ambiguity, data quality, bias and fairness, map out students' implementation barriers, and the pedagogical effectiveness of manual data annotation as a teaching concept that highlights subjectivity of data annotation.

Self-reported familiarity with the course content increased substantially across all concepts. Most of the students recognised that personal interpretation affects annotations. The students rated the activity as more effective than traditional lectures in understanding bias. They were also motivated to learn more. However, emotional discomfort from viewing medical images was the main drawback. Many students still requested clearer guidelines to reduce disagreement. This suggests that they had not yet internalised that disagreement grounded in different perspectives is a feature, not a bug.

Manual data annotations serve as effective pedagogical hints, teaching students that human judgement influences model behaviour. They also show that disagreement reflects domain complexity, not just noise. Our main takeaways for future iterations of this or similar courses are: ensure sufficient interpretive ambiguity, reduce repetitive workload, mitigate emotional discomfort, and explicitly frame disagreement as a learning opportunity rather than a problem to solve.
    \end{abstract}
    \vspace{0.5em}
    \vspace{1em}

    \begin{disclaimerbox}
    \noindent\textbf{Disclaimer:} This is a working paper, and represents research in progress. We welcome contributions from the community. For comments or questions please email us at \href{mailto:ralf.raumanns@fontys.nl}{ralf.raumanns@fontys.nl} or \href{mailto:theresa.elstner@uni-kassel.de}{theresa.elstner@uni-kassel.de}.
    \end{disclaimerbox}
  \end{@twocolumnfalse}
]

\section{Introduction}\label{Introduction}
    Machine learning (ML) and deep learning (DL) are growing rapidly, thus becoming a core element of the educational curriculum of universities. A core task of universities is to bring together research and education. Findings and insights from science are translated into concrete educational experiences, giving students access to current knowledge while learning how research is conducted. In a traditional approach, converting ML and DL research into a curriculum involves lectures and practical exercises that focus on the technical implementation of models using pre-labelled datasets, rather than on the process of generating and curating them. One of the aspects that is lost in this approach is that students develop a feel for the data: because they do not go through the often non-trivial annotation process themselves, they miss its inherent subjectivity and complexity. By subjectivity, we mean that annotation choices depend on personal interpretation, not just on a single objective ground truth. In many ML tasks the label is not uniquely determined but influenced by human judgement, where (even expert) annotators may legitimately disagree.

\begin{figure}[t]
    \centering
    \begin{minipage}{0.4\textwidth}
        \centering
        \begin{subfigure}[b]{0.45\textwidth}
            \centering
            \includegraphics[width=\textwidth, height=3cm, keepaspectratio]{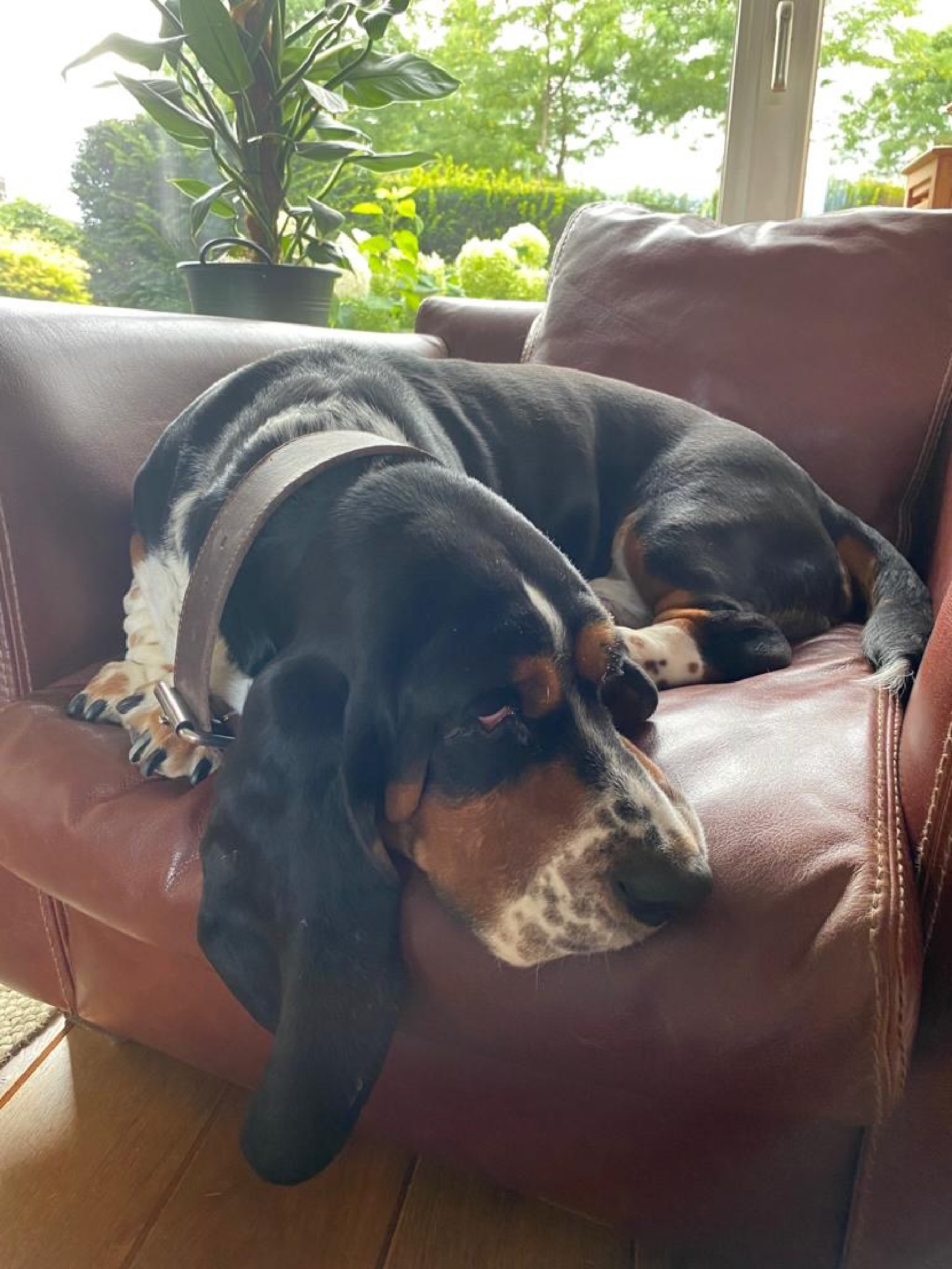}
            \caption{}
        \end{subfigure}
        \hfill
        \begin{subfigure}[b]{0.45\textwidth}
            \centering
            \includegraphics[width=\textwidth, height=4cm, keepaspectratio]{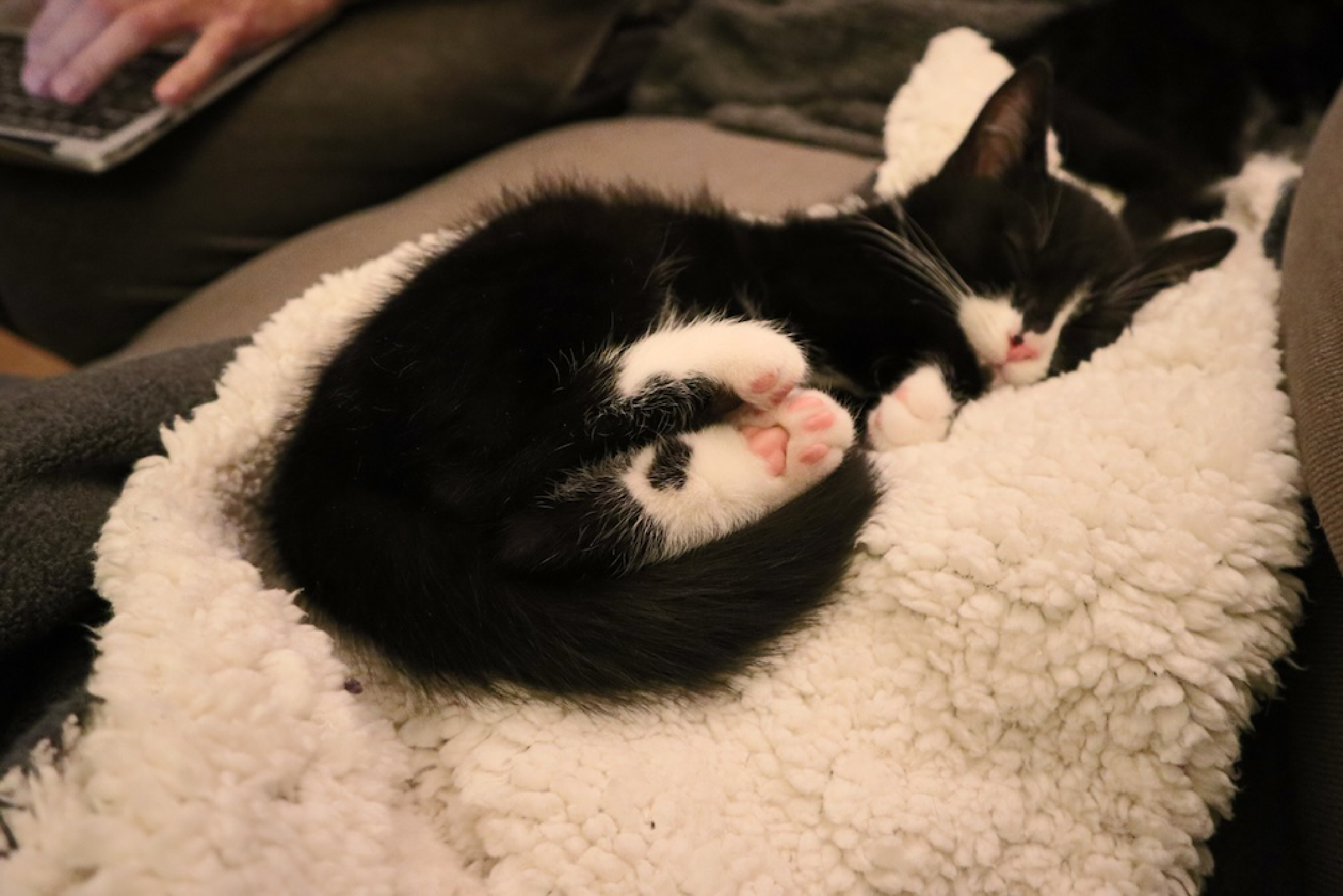}
            \caption{}
        \end{subfigure}
        \vspace{0.5em}
        \begin{subfigure}[b]{0.45\textwidth}
            \centering
            \includegraphics[width=\textwidth, height=4cm, keepaspectratio]{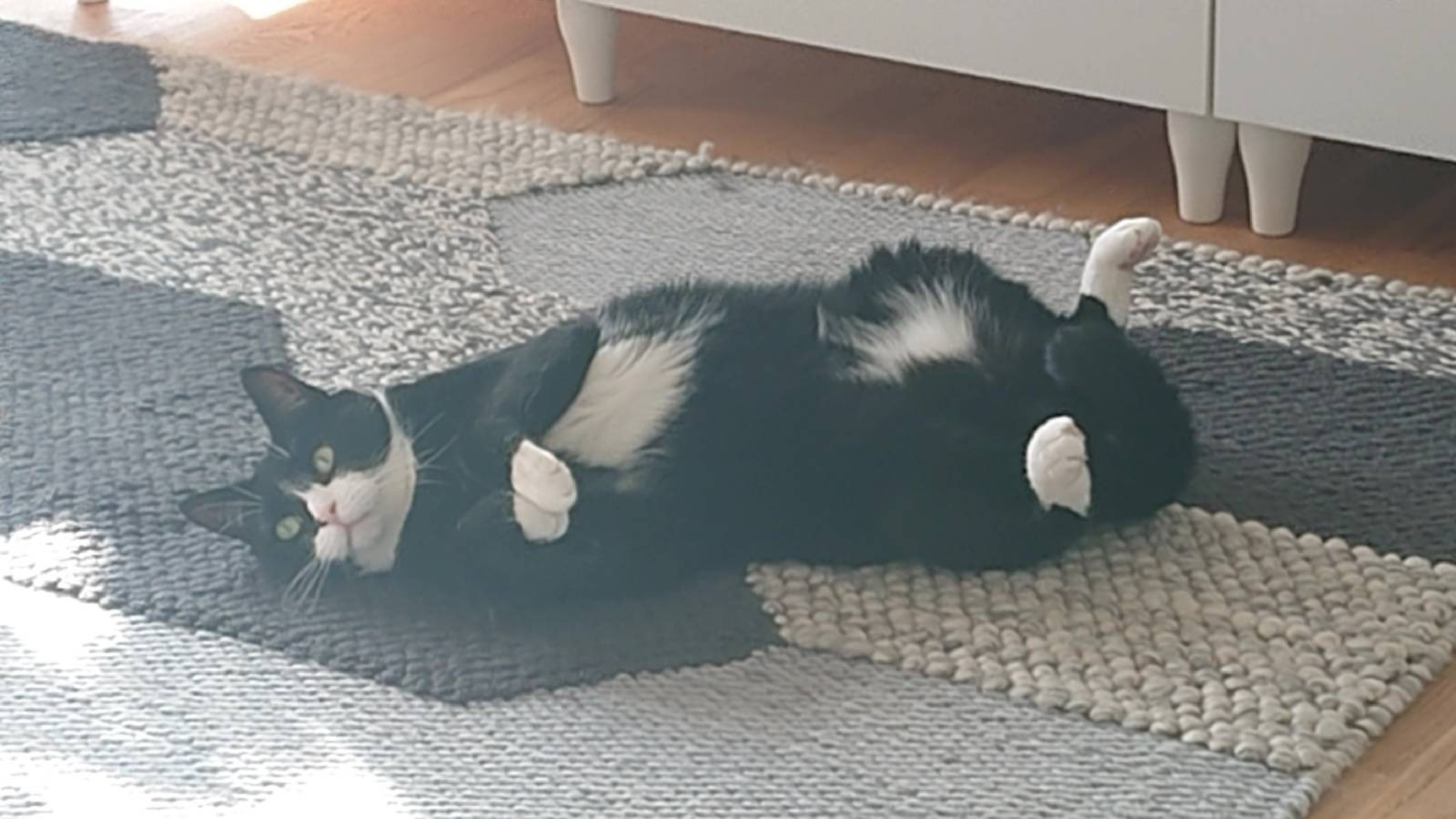}
            \caption{}
        \end{subfigure}
        \hfill
        \begin{subfigure}[b]{0.45\textwidth}
            \centering
            \includegraphics[width=\textwidth, height=3cm, keepaspectratio]{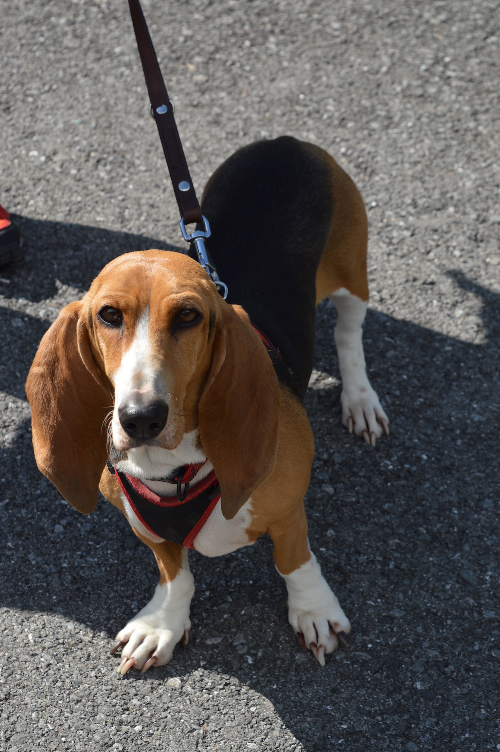}
            \caption{}
        \end{subfigure}
    \end{minipage}
    \caption{Demonstrating the impact of task design on annotation consistency. The same four images are subjected to two different tasks: \textbf{Task 1} is an objective classification ('Cat' vs. 'Dog') that typically yields unanimous agreement. In contrast, \textbf{Task 2} asks for a subjective assessment of activity level ('Calm', 'Neutral', 'Playful'), which naturally introduces interpretive ambiguity and disagreement among annotators.}
    \label{fig:task_examples}
    \end{figure}

Figure \ref{fig:task_examples} shows this: Task 1 (Cat vs. Dog) yields clear agreement, while Task 2 (Calm, Neutral, Playful) creates natural disagreement. By complexity, we mean the challenges during annotation: such as dealing with ambiguous cases, handling different opinions, and understanding how personal bias affects training data.

This paper focuses on data and annotation because the process of gathering training data is an important source of bias, separate from model architecture or algorithmic choices. While models can introduce bias through their design, our work addresses the earlier stage where human judgement shapes what the model learns.

We propose a pedagogical shift: students create \textit{hints} themselves instead of using a pre-labelled dataset. By \textit{hints}, we refer to annotations of visual features in an image, such as the border irregularity or colour variation of a skin lesion, that serve as interpretable signals for a model rather than as definitive ground truth. Pre-labelled datasets offer practical advantages such as time efficiency and reduced workload, but they hide the annotation process from students. Creating hints serves two goals: it teaches students about the relation between human judgement and model behaviour, while generating training data. Secondly, it engages students directly with research and the nuances of labelling, thus deepening their understanding of how human judgement shapes the data underlying model performance and interpretability.

When we subsequently integrated a comparable annotation task into our data science courses, we observed an unintended pedagogical effect: students began to critically question annotation protocols, discuss disagreements with peers, and recognise how their interpretive choices could propagate into model training. This observation suggests that data annotation may serve not only as a data collection mechanism but also as a vehicle for developing students' awareness of subjectivity, bias, and fairness in AI. The present study investigates this potential in a systematic way.

We have developed an educational activity and implemented it as a case study at two universities. In both institutions, students were given a real-life dataset with a diverse collection of skin lesions. Students actively annotated these images for visual features such as hair coverage. Working in groups, they applied a 3-point scale (none, some, a lot) to hundreds of images, documented their annotation decisions, and compared their labels with peers to analyse disagreement. The activity emphasises critical reflection on the collected data, combining elements of active learning, problem-based learning, and shared tasks.

We surveyed the students who took the data science course that contained the annotation task.

Through this case study, we explore the answers to two research questions.
\begin{enumerate}
    \item To what extent does annotating data in groups affect students’ perception of data quality and its effect on bias and fairness in AI models? (RQ1)
    \item From the perspective of students and educators, what are the perceived pedagogical strengths and limitations of using data annotation as an educational task? (RQ2)
\end{enumerate}

Drawing on the insights from the survey, we propose recommendations for designing annotation assignments that involve students directly in the research process. Our primary audience is data science and ML educators. Our contributions are:
\begin{itemize}
    \item A practical educational intervention using skin lesion annotation to teach responsible AI concepts
    \item Exploratory insights (from surveys and reflections) on the pedagogical value of this approach across two institutions
    \item Practical takeaways for designing annotation-based assignments that connect human judgement to model behaviour
\end{itemize}

\section{Related work}\label{Relatedwork}
    In this section, we review (i) classical agreement metrics that assume ground truth consensus, (ii) emerging approaches that model annotator subjectivity, and (iii) the gap between these methodological developments and their treatment in computer science education.

\subsection{Data annotation as a research problem: from ground truth to subjectivity}
\label{sec:agreement-disagreement-paradigm}

Understanding how data labels are created is important to understanding what ML models learn. The nature of annotation itself has undergone a quiet transition: disagreement is increasingly treated not as noise but as meaningful information. This section reviews this shift, from the agreement paradigm to recent approaches that accept subjectivity, and discusses how these insights have yet to enter education.

\paragraph{Traditional view on data annotation: the agreement paradigm}\label{agreement_paradigm}
Data annotation has traditionally been grounded in the assumption of a single latent ground truth. Under this view, disagreement between annotators is treated as noise, and annotation quality is commonly evaluated using inter-annotator agreement measures such as Krippendorff's alpha~\cite{krippendorff:1970}, Fleiss's kappa~\cite{fleiss:1971}, Cohen's kappa~\cite{cohen:1960}, and Kendall's tau~\cite{kendall:1938}. These measures remain widely used in ML research, particularly in supervised learning settings where human annotations serve as gold-standard labels.

\paragraph{Shift to subjectivity of data annotation: the disagreement paradigm}
More recently, the assumption of a single latent ground truth has been challenged. A growing body of work argues (\cite{roettger:2022, uma:2022, sandri:2023, aroyo:2015, dumitrache:2018, cheplygina:2018, liao:2024, plank:2022}) that disagreement between annotators can reflect a meaningful variation in interpretation rather than measurement error. In this view, multiple perspectives may coexist for inherently subjective tasks,
and annotation should not necessarily be reduced to a single aggregated label.

R{\"o}ttger et al. \cite{roettger:2022} formalise this distinction through descriptive versus prescriptive annotation paradigms. The prescriptive view assumes a single correct label, while the descriptive view allows multiple valid interpretations of the same data.
Similarly, Uma et al.~\cite{uma:2022} show that preserving disagreement can even improve downstream performance compared to relying solely on aggregated gold labels. 

Several works further investigate the structure and usefulness of disagreement. Sandri et al.~\cite{sandri:2023} provide a taxonomy of the causes of disagreement among annotators, distinguishing between noise-driven sources (sloppy annotation, ambiguity, and missing information) and genuinely subjective sources.
Aroyo and Welty~\cite{aroyo:2015} argue that disagreement itself can be informative and propose the Crowd Truth framework, which models labels as distributions over interpretations rather than as single ground truth values. Building on this idea, Dumitrache et al.~\cite{dumitrache:2018} develop metrics to quantify ambiguity across annotators, data, and annotations.

Empirical evidence in specific domains supports this perspective. Cheplygina and Pluim~\cite{cheplygina:2018} show that disagreement in medical image annotation can carry informative signals towards classification. More recently, the Preference-involved Annotation Distribution Learning (PADL) framework~\cite{liao:2024} has expanded this direction by training annotator-specific image segmentation models based on annotations by multiple medical professionals.
Calling out to the research community, Plank~\cite{plank:2022} argues for publishing and leveraging non-aggregated annotations in order to devise new directions for ML.

Together, these works reflect a shift in the field: from treating annotation disagreement as a problem to be eliminated, toward treating it as a signal that may encode meaningful structure about subjective tasks.

If research sees disagreement as meaningful, why does this insight rarely reach the classroom? In typical ML courses, students work with pre-labelled benchmark datasets, so the question of how labels were created, and how annotators may disagree, is often not addressed at all. Our study explores whether engaging students in annotation can connect modern theory to classroom practice.

\subsection{Data annotation in teaching: persistence of the agreement paradigm} 
\label{sec:data-annotation-in-teaching}

In several educational settings, annotation exercises involve computing inter-annotator agreement measures such as Cohen's kappa or Krippendorff's alpha. For example, Fr{\"o}be et al.~\cite{froebe:2024} describe an information retrieval course in which students construct test collections and evaluate annotation quality using agreement metrics. Crucially, however, these courses typically do not include an assignment to merge or reconcile disagreements afterward, leaving the implicit message that consensus is the endpoint.

Similarly, even recent NLP teaching materials often rely on pre-labelled datasets and focus on model building and evaluation rather than on the annotation process itself (\cite{parde:2024},\cite{nikishina:2024},\cite{assenmacher:2024},\cite{biester:2024},\cite{helcl:2024},\cite{lee:2024},\cite{anderson:2024},\cite{ginn:2024},\cite{prasad:2024},\cite{micluta:2026}). Although some courses incorporate manual annotation tasks, these are typically designed for unambiguous labelling problems such as part-of-speech tagging~\cite{barteld:2017}\cite{joshi:2024}, where subjectivity is limited.

Even when subjective tasks are included, educational framing often remains consensus-oriented, following the traditional agreement paradigm. Cignarella et al.~\cite{cignarella:2024} and Hou~\cite{hou:2024} describe annotation exercises in which student disagreement is addressed through discussion aimed at resolving differences and reaching agreement, rather than exploring disagreement as a meaningful signal.

Beyond computer science, educational approaches similarly emphasise annotation as a means of supporting domain learning or critical engagement with subject matter rather than as a site of epistemic reflection on label subjectivity~\cite{gholiagha:2025} \cite{braz-sousa:2024}.

Overall, while annotation is increasingly used as a pedagogical tool, its epistemological framework in education largely reflects the traditional agreement paradigm identified in research.

In this study we explore this gap by examining whether engaging students in manual labeling helps them recognise disagreement as meaningful rather than as error.

\subsection{Linking research and teaching}

\paragraph{Didactics of research-oriented teaching} 

In a conventional curriculum, theoretical concepts are taught and reinforced through practical assignments, reflecting a \emph{transmission perspective} on teaching \cite{pratt_five_1998}. Content-driven approaches assume stable knowledge, which fast-moving fields like ML and data science no longer offer. As a result, demand is rising for curricula that more directly incorporate current research \cite{Healey2005-cj}, enabling students to engage with recent findings and challenges.

Elstner et al. \cite{Elstner2024-ud} propose the integration of research to teaching through \textit{shared task} activities. In these tasks, students collaborate in groups to tackle a scientific problem, with the emphasis on the research content rather than competition. The authors ground their instructional design in active‑learning theory~\cite{felder:2009}, positioning shared tasks as a specialised form of \textit{project-based learning} (PBL).
Raschka \cite{Raschka2021-qi} further illustrates the efficacy of PBL in ML education. Students formed teams, developed independent project proposals, and spent the semester iteratively building and evaluating their models. At the end of the semester, each group presented its work to peers. An anonymous post‑course survey indicated high levels of student satisfaction, and the author reported that PBL also fostered increased interaction and cooperation among participants.

PBL increases learners' interest in the material \cite{Pease2011-rd}. Our work brings together two related ideas: (1) linking current research more closely to coursework, and (2) using active‑learning methods, like shared task activities and PBL, to provide students with hands-on experience in applied data science. 

Schiendorfer et al. \cite{Schiendorfer2020-pk} observe that computer science and software engineering students tend to jump straight to building a solution rather than conducting empirical investigations first. To foster a more empirical mindset, they emphasise proper experimental work when teaching ML to these students.

By placing real research problems in the classroom, we combine the motivation that PBL provides with the benefit of exposing students to up‑to‑date scientific work, thereby positioning our study in the field of research‑driven, student-centred education.

\paragraph{Modern annotation theory and education}
Taken together, ML research has increasingly emphasised that annotation disagreement can encode meaningful structured variation rather than noise. In contrast, educational practice continues to rely predominantly on agreement-based evaluation frameworks, implicitly reinforcing a single-ground-truth view of annotation.

Prior work has demonstrated that annotations collected from non-experts, including students, can serve as effective training signals for ML models (\cite{raumannsr-melba}, \cite{raumanns2020multi}).
The authors trained a multi-task DL model on student-generated annotations of visual characteristics such as asymmetry, used in dermatological practice as part of the ABCD rule \cite{Stolz1994-px}, showing that crowdsourced features can improve both diagnostic performance and interpretability. However, in these studies, the focus lay on using the resulting data for model training rather than on systematically investigating the instructional value of the annotation activity itself.

To our knowledge, no prior work in data science or computer science education has systematically investigated how students conceptualise annotation subjectivity when not explicitly guided toward either the traditional or modern perspective. In particular, it remains unclear whether students naturally adopt an agreement-orientated interpretation of the annotation or whether they independently recognise disagreement as potentially informative.

This gap motivates our exploratory study, which examines student perceptions of annotation subjectivity as a first step toward addressing these questions (see contributions in the Introduction).

\section{Methods}\label{Methods}
    \subsection{Educational context}
    Two educational courses formed the basis of this study, each representing a distinct approach to teaching data science, delivered by several of the authors. The first course, \textbf{Data Mining}, was offered during the fourth semester at Fontys University of Applied Sciences in Venlo. In this 5-ECTS course, we introduced the core concepts of ML and data-driven decision-making, structured around IBM's data science methodology. The second course, \textbf{Projects in Data Science}, took place at the IT University of Copenhagen. In this 7.5-ECTS course, we used a project-based learning approach to guide students through the entire data science pipeline, from problem identification to solution implementation and communication. In addition, an annotation task was designed and implemented in both courses. In this activity, participants worked in groups over a two-week period to annotate skin lesion data.

We chose medical image annotation because it is naturally subjective while still being connected to real clinical decisions. Importantly, students do not label the diagnosis directly; instead, they annotate visual features that may help determine it. Connecting the annotation task to real clinical decisions helps students understand that their choices matter. This makes the lesson about subjectivity and bias concrete rather than abstract. Recent work in medical imaging \cite{liao:2024} states that ``medical image annotation is highly subjective, leading to inevitable annotation biases.'' Scoring hair coverage in skin lesion images works especially well for our purpose. Unlike objective tasks such as distinguishing cats and dogs, judging how much hair is visible requires interpreting features such as thickness and density in different areas of the same lesion. The same image can reasonably be judged as having ``some hair'' by one person and as having ``a lot'' by another. Medical images can also trigger emotions, making the task feel more real, but can also cause discomfort. Together, these factors create a task in which students naturally disagree, letting them see for themselves how human judgement shapes training data.

\begin{table*}[htbp]
  \centering
  \small
  \caption{Overview of the two courses in which the annotation activity was implemented}
  \label{tab:courses}
  \begin{tabular}{@{}lp{5.5cm}p{5.5cm}@{}}
    \toprule
    \textbf{Characteristic} & \textbf{Data Mining (\DAMI)} & \textbf{Projects in Data Science (\PDS)} \\
    \midrule
    Institution & Fontys University of Applied Sciences, Venlo (The Netherlands) & IT University of Copenhagen (Denmark) \\
    
    Period & Fourth semester, 2024-2026 & Second semester, 2025-2026 \\
    
    Credits & 5 ECTS & 7.5 ECTS \\
    
    Focus & Core concepts of ML and data-driven decision making & Complete data science project pipeline on a domain-specific problem: image processing for segmenting and classifying skin lesions (PAD-UFES-20 dataset \cite{Pacheco2020-vz})  \\
    
    Key topics & Data exploration, dimensionality reduction, supervised learning, and reinforcement learning, all organised according to IBM's data science methodology & Image representation and processing, annotation and agreement analysis, segmentation evaluation, feature representation, classifiers, evaluation, and handling class imbalance. \\
    
    Pedagogy & Regular weekly theory lectures and hands-on sessions, supported by short assessments, incorporating both individual and group activities & Project-oriented instruction combining weekly lectures (2×45 min) and hands-on sessions (2×45 min), featuring ungraded practice tasks with published solutions; includes collaborative group work \\
    
    Assessment & Written exam and practical assignments (both mandatory) & Project-oriented format; assessment through a group submission and an oral examination (team presentation followed by individual questioning) \\
    
    Prerequisites & Prior knowledge of Python and data analysis & Background coursework in data science, linear algebra, and probability, with simultaneous enrollment in second-semester classes \\
    
    Tools \& platforms & Python, Scikit-Learn, TestVision & Python, Scikit-Learn, GitHub, Overleaf \\
    
    Learning materials & Course slides and the textbook \emph{Hands-On Machine Learning with Scikit-Learn and TensorFlow} \cite{Geron2022-uo} & Course slides and exercise notebooks; selected research papers (optional reading)\\
    \bottomrule
  \end{tabular}
\end{table*}







Table~\ref{tab:courses} provides a comparative overview of the key characteristics of both courses, including their institutional context, curriculum focus, pedagogical strategies, assessment methods and required tools.

\subsection{Intervention design}\label{interveniton_design}
    We integrated case-based learning activities into each participating course. These courses already existed in the regular curriculum and focused on practical skills in ML and data science. We used these existing activities to investigate the pedagogical impact of data annotation, with an emphasis on ambiguity, subjectivity, and bias in data interpretation. The annotation task was divided into two parts: an individual assignment and a group assignment.

\subsubsection{Individual assignment: introduction to data annotation}
    In the first assignment, students worked with existing datasets of skin lesion images. Both institutions initially used annotated images from the ISIC archive (\cite{Gutman2016-lf,Codella2017-rd,Codella2019-cn,Tschandl2018-sz,Combalia2019-jj,Veronica2021patient}); in later iterations, ITU switched to the PAD-UFES-20 dataset of clinical smartphone images \cite{Pacheco2020-vz}, while Fontys continued with ISIC. The students explored this dataset using a Jupyter notebook, which guided them through exercises involving data exploration, agreement between the annotators, and critical reflection on annotation variability. This activity emphasised the complexity of subjective data annotation and the importance of consistency and transparency in annotation practices.

\subsubsection{Group assignment: hands-on data annotation}
    The second assignment was a collaborative group task in which each group was assigned a unique set of skin lesion images: students from ITU received 100 images, while students from Fontys received 200 images. The Fontys students annotated the amount of visible hair in each image using the open-source tool Label Studio \cite{labelstudio}. The ITU students recorded their scores in a standardised CSV template. In both cases, the students annotated the amount of visible hair in each image on a 3-point ordinal scale (0 = none, 1 = some, 2 = a lot). The groups consisted of five students at ITU and at most four students at Fontys. Each group was required to annotate at least 100 images, with all members annotating the same subset to enable comparison and analysis of inter-annotator agreement. Within-group agreement was quantified using Fleiss's $\kappa$; at Fontys this calculation was part of the students' assignment, at ITU it was performed by the authors. Students were encouraged to document their annotation decisions and reasoning, facilitating reflection on subjectivity and potential biases in visual interpretation.\\

    Both assignments were integrated into the standard course design at each University and were consistent with the corresponding course’s learning objectives. Depending on the course format, students carried out the tasks either individually or in small groups. Instructors and teaching assistants provided formative feedback during practical sessions, and students were encouraged to reflect on their annotation strategies and challenges. These activities aimed to develop students’ critical thinking, data literacy, and ethical awareness in the context of ML.

\begin{figure}[t]
\centering
\begin{minipage}{0.4\textwidth}
    \centering
    \begin{subfigure}[b]{0.4\textwidth}
        \includegraphics[width=\textwidth]{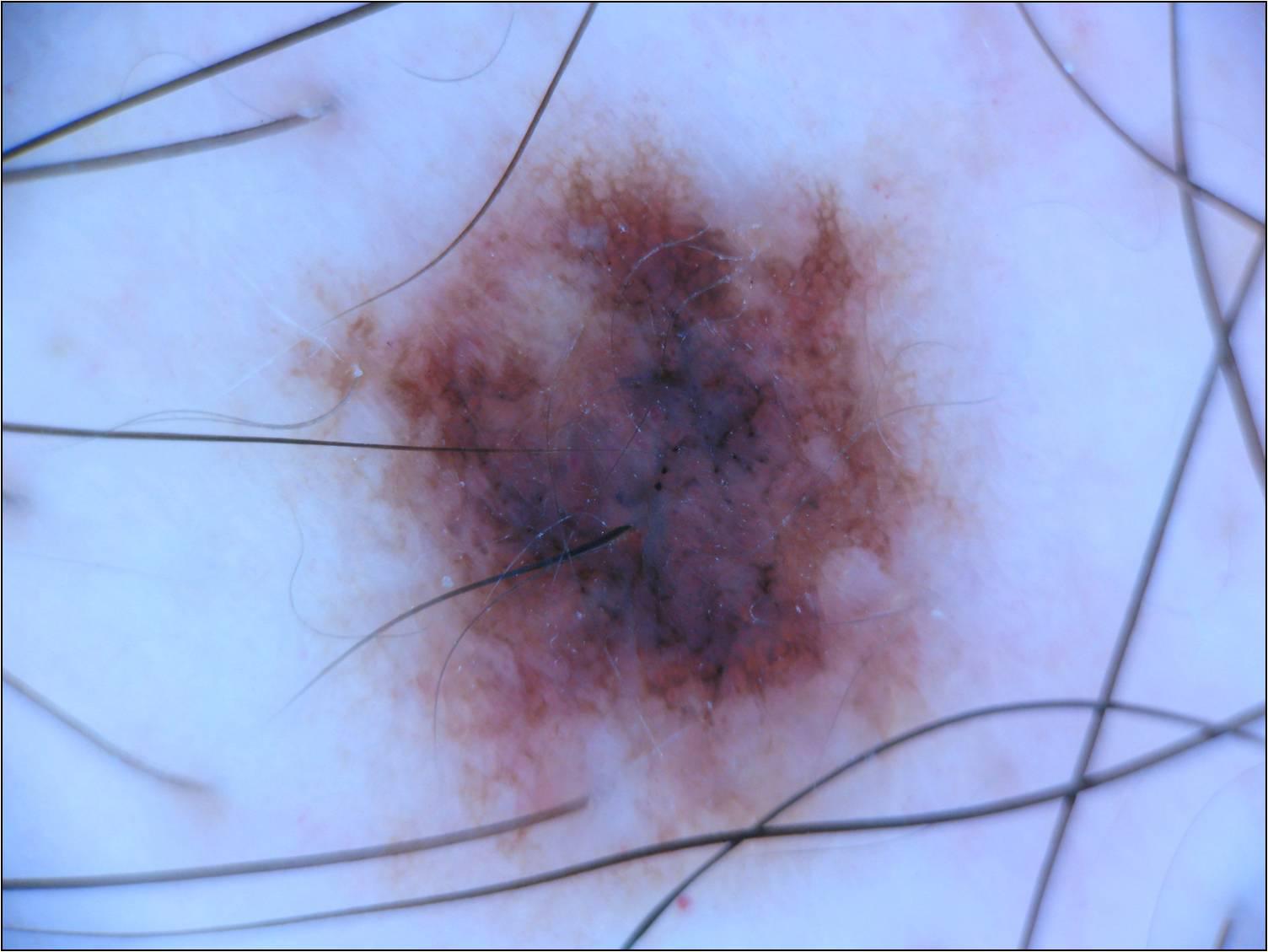}
        \caption{}
    \end{subfigure}
    \hfill
    \begin{subfigure}[b]{0.4\textwidth}
        \includegraphics[width=\textwidth]{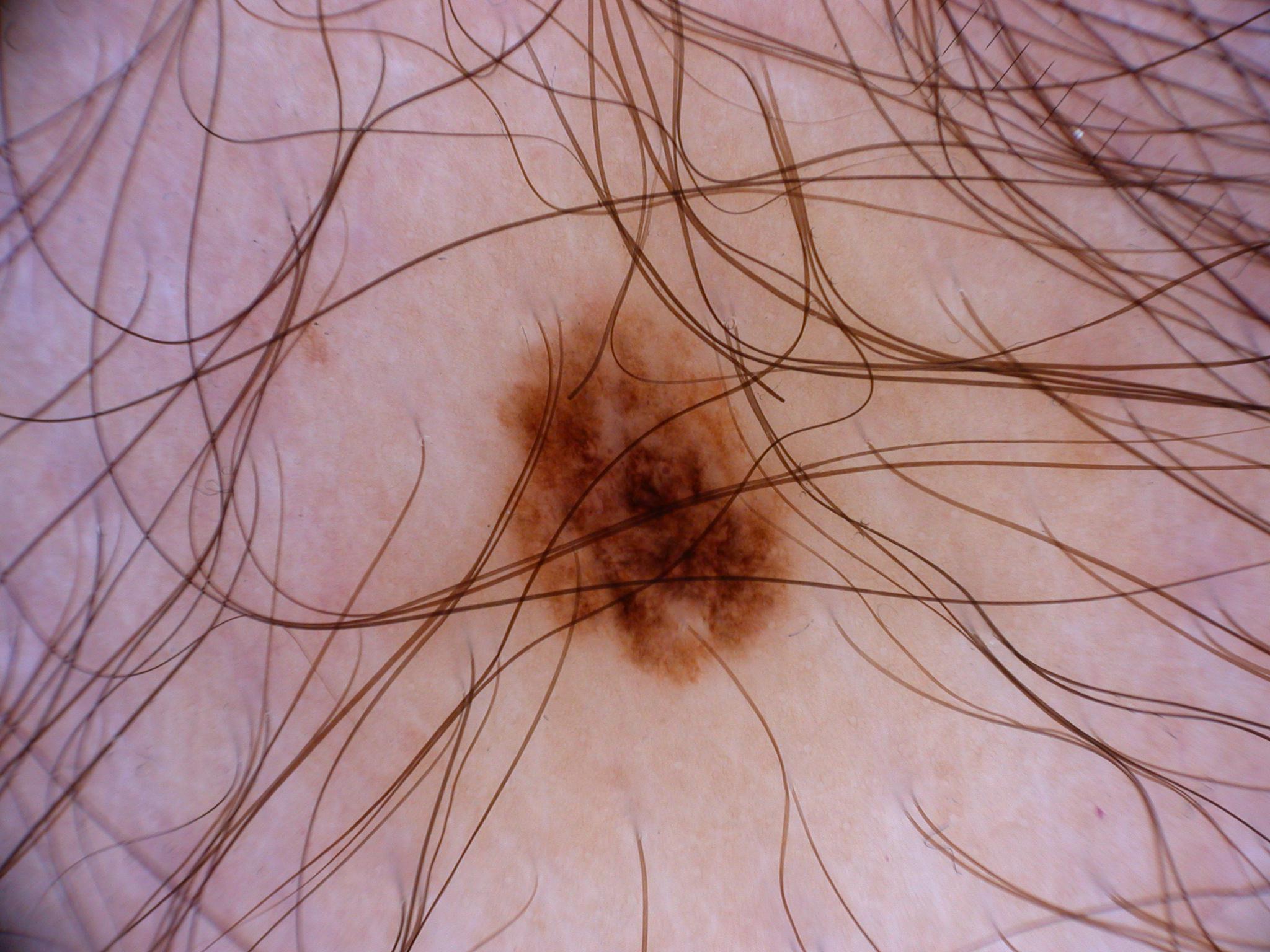}
        \caption{}
    \end{subfigure}
    \vfill
    \begin{subfigure}[b]{0.4\textwidth}
        \includegraphics[width=\textwidth]{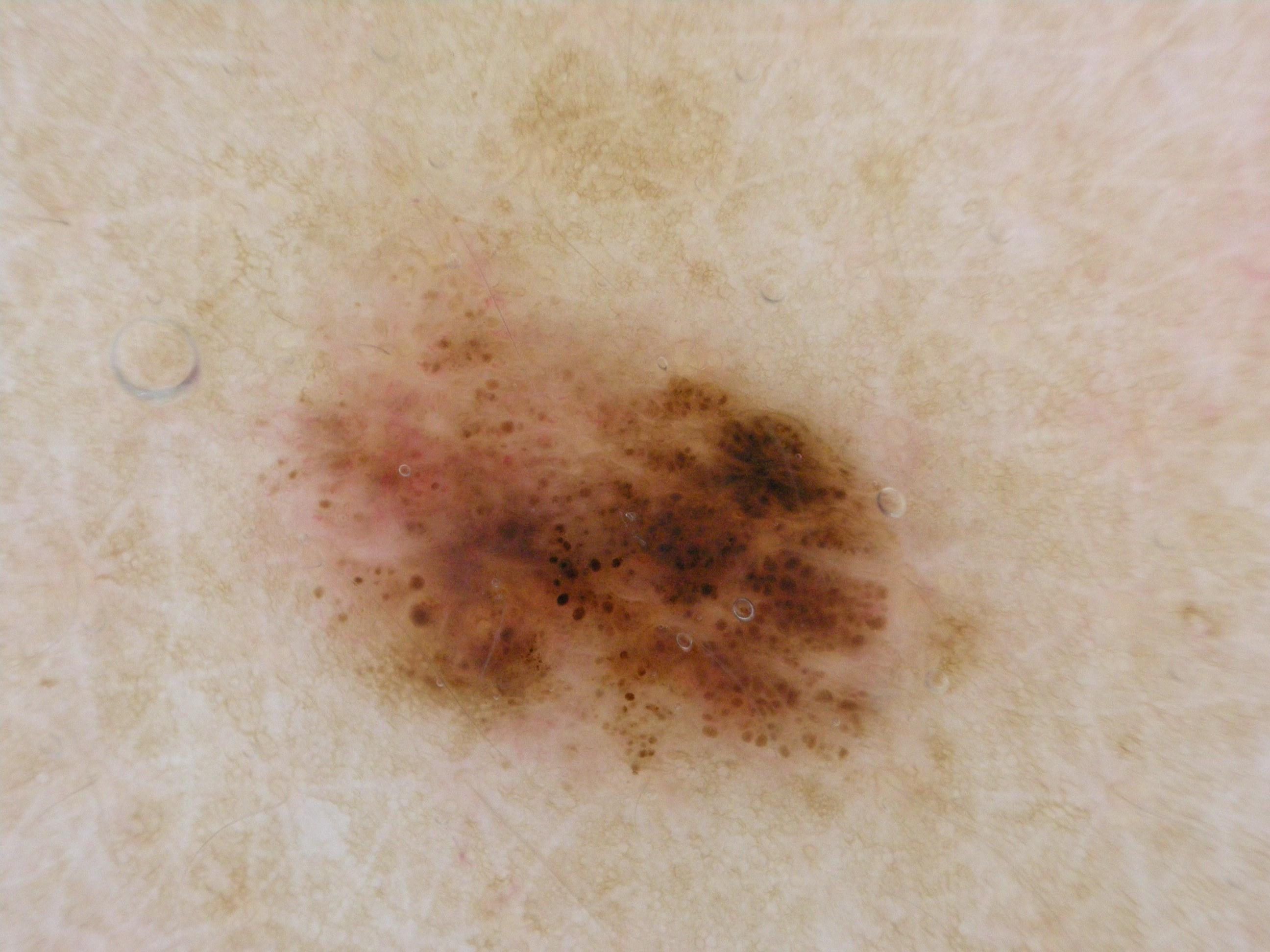}
        \caption{}
    \end{subfigure}
    \hfill
    \begin{subfigure}[b]{0.4\textwidth}
        \includegraphics[width=\textwidth]{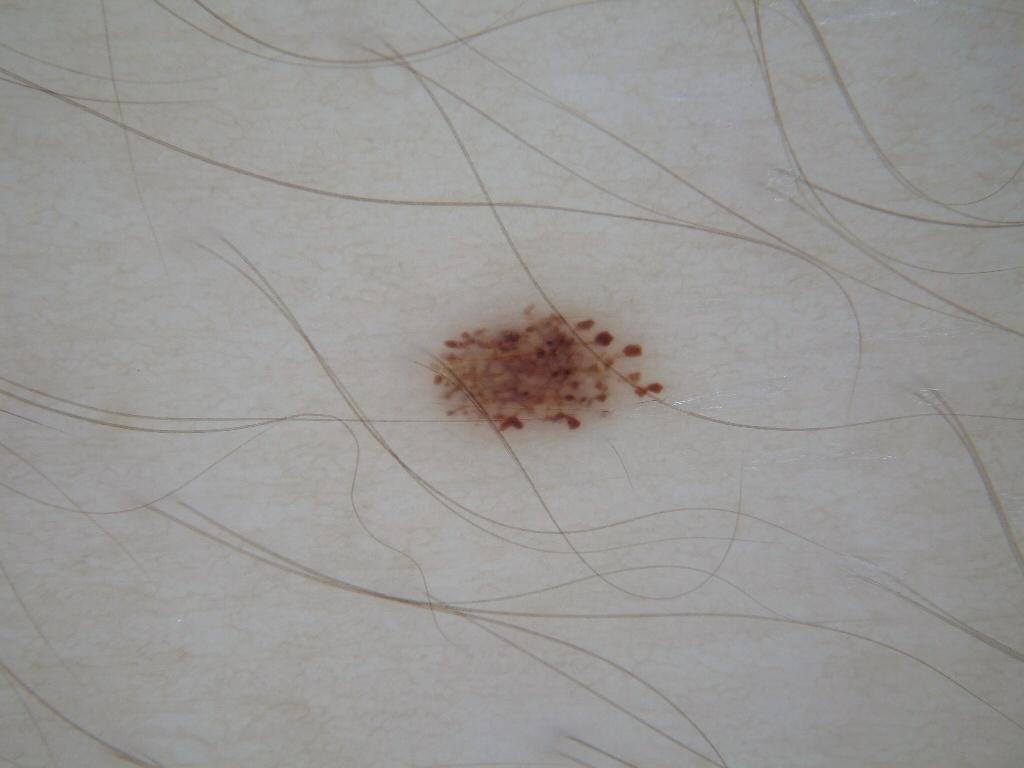}
        \caption{}
    \end{subfigure}
\end{minipage}
\caption{Examples of four distinct skin lesions demonstrating varying degrees of hair coverage. For an experience of subjectivity, we invite readers to evaluate the hair density for each lesion: 0 no visible hair, 1 some hair presence, and 2 a lot of hair.}
\label{fig:skin_lesions_examples}
\end{figure}

\subsection{Survey instrument}
    We implemented a single online survey after the students completed the annotation task. The survey consisted of three sections: (1) a retrospective pre- / post-assessment, (2) closed-ended, and (3) open-ended questions inviting students to elaborate on their experiences. We hosted the survey on SoSci Survey \cite{Leiner-nm} and collected responses between 27 May 2026 and 24 June 2026.

\begin{table*}[htbp]
  \centering
  \caption{Overview of the five constructs, definitions, and sample items}
  \label{tab:constructs}
  \begin{tabular}{@{}p{3cm}p{4cm}cp{5cm}@{}}
    \toprule
    \textbf{Construct} & \textbf{Operational Definition} & \textbf{\# Items} & \textbf{Sample Items} \\
    \midrule
    Annotation ambiguity & The systematic disagreement among annotators regarding the annotation of the same instance, arising from inherent ambiguity, complexity, or subjectivity in the data rather than from annotation errors. & 4 & ``The annotations produced by my group were consistent with each other'' \\
    Data quality & The understanding how subjective interpretation and annotation challenges influence training data. & 3 & ``Doing the annotation task made me more aware of the challenges
in creating training data.'' \\
    Bias and fairness & The understanding how subjective disagreements and confounding visual cues in data lead to unreliable AI predictions & 5 & ``The annotation task gave me insight into why AI models can
produce unreliable classifications.'' \\
    Implementation barriers & The obstacles related to instructions, tools, emotional comfort, or self-confidence that impede effective participation in the annotation task & 6 & ``I felt confident in my ability to accurately label the skin lesion images'' \\
    Pedagogical effectiveness & The extent to which the annotation task increases student motivation, interest, and understanding of key concepts such as subjectivity, disagreement, and how human judgment influences model behaviour. & 6 & ``What is the most enjoyable part of the task?'' \\
    \bottomrule
  \end{tabular}
\end{table*}

\subsubsection{Survey design}
    The first section of the survey focused on retrospective pre-task and post-task familiarity measures. We asked participants to reflect on their understanding prior to starting the annotation task across four core concepts: subjectivity in data annotation, inter-annotator agreement measures, bias in ML models, and data quality assessment. This allowed us to compare self-reported knowledge before and after the intervention on a 5-point scale ranging from ``never heard of it'' to ``fully understood it''.
    
    The second section aimed to assess construct-related perceptions regarding the annotation task experience. Following an iterative development process, we translated five constructs into measurable survey items covering annotation ambiguity, data quality, bias and fairness, implementation barriers, and pedagogical effectiveness (See Table \ref{tab:constructs}). All items used a 7-point scale ranging from ``strongly disagree'' to ``strongly agree'', enabling systematic measurement of student opinions. We developed multiple items per construct to minimise measurement error.
    The third section captured qualitative reflections through open-ended questions. We gave respondents the opportunity to elaborate on their views in their own words, including what they found most significant about the annotation process, suggestions for improving the task, least enjoyable parts of the task, most enjoyable parts, and what they found interesting about skin lesions as a topic.

    Participants were informed of the purpose of the study and that participation in the survey was voluntary. At Fontys, where the annotation task was a mandatory course component, students were informed that completing the annotation task was required for their coursework, but that participating in the research survey was not. At ITU, the annotation task was a mandatory component of the course, while the accompanying survey was voluntary. We stored all anonymous responses on the SoSci Survey platform, which is GDPR-compliant. Data were not linked to student names or student numbers, and participation in the survey did not affect course grades. 
    
    The complete version of the survey is included in \textbf{Appendix A}.

\subsubsection{Survey rollout and response collection}
After the annotation task was completed, we sent the survey in different ways for each university. At Fontys, we asked the 2025–2026 students to do the survey during class. We emailed the 2024–2025 students at Fontys after their course ended. At ITU, we posted flyers on campus. We retained only fully completed responses for analysis. 

\subsection{Data analysis}
    \subsubsection{Quantitative analysis}  
    We analysed post-survey data using descriptive statistics (frequencies, percentages, and Likert-scale distributions) to summarise responses and calculate proportions for each construct. These variables addressed our core constructs: annotation ambiguity, data quality, bias and fairness, implementation barriers, and pedagogical effectiveness. These constructs were selected to align with both our research questions (RQ1 and RQ2) and the learning persona trajectory. Together, they trace the cognitive and learning path we expect students to follow: from encountering ambiguity during the annotation activity, to recognising its implications for data quality, to connecting it to AI bias. Annotation ambiguity and data quality capture the immediate annotation experience. Annotation ambiguity measures whether students perceive disagreement as a natural consequence of subjective interpretation, while the data quality assesses whether they transfer this realisation to a broader understanding of how training data is constructed. These constructs directly address RQ1, which asks how group annotation affects perceptions of data quality. Bias and fairness extends this by measuring whether students connect their annotation experience to the downstream consequences for AI models, which is central to RQ1. Implementation barriers and pedagogical effectiveness address RQ2 by capturing the practical and educational sides of the intervention.
    
    Additionally, we evaluated agreement among annotators within each student group using Fleiss’s $\kappa$ (suitable for multiple raters, unlike Cohen's $\kappa$). This allowed us to examine whether participants’ subjective impressions of within-group agreement aligned with the objectively calculated statistics. For Fontys groups, we extracted $\kappa$ results from the Jupyter Notebook files submitted by each group via the Canvas learning management system. For ITU groups, we downloaded the annotation files \cite{projectDataScience2026ExamTemplate} and calculated $\kappa$ ourselves.

\subsubsection{Qualitative analysis} 
    We followed the six-step thematic analysis framework of Braun and Clarke \cite{Braun2006-zq}: familiarisation with the data, generating initial codes, searching for themes, reviewing themes, defining and naming themes, and producing the report. As an adaptation, we merged the searching and reviewing of themes into a single collaborative consensus step, in which two researchers independently coded the responses and subsequently reconciled their codes through discussion. 
    
    We exported all data from open-ended questions in Appendix A (AF1, IB5, IB6, PE5, PE6). We (authors TE and RR) proceeded as follows: First, we familiarised ourselves with the data by reading all responses repeatedly. Next, we independently performed theme modelling for each question. This allowed us to generate initial themes that captured meaningful patterns. We then discussed these results to reach a consensus on the final list of themes aiming at full coverage of individual answers under unified level abstraction of themes. Next, both authors independently annotated each response with its corresponding themes using Doccano~\cite{Doccano}. We measured inter-author agreement and obtained a relaxed F1 score of $0.71$, which counts overlapping text spans as matches only when their labels are identical. Based on the final annotations, we counted theme frequencies per open-ended question to quantify their prevalence across the constructs. We report the average frequency per theme ($\bar{n}$), calculated as the total count divided by the number of annotators (two).

\subsection{Analytical framework: the learning persona}
    To systematically evaluate student responses, we used an analytical framework centered on the learning persona, a theoretical construct derived from user‑centred design principles \cite{Pruitt2010-by}. The persona represents what students should ideally understand after completing the annotation task.

The framework rests on three interconnected assumptions about how learners process subjective information. First, students typically approach annotation tasks with a traditional mindset that assumes a single objective ground truth, treating disagreement as error. The educational goal is to shift them toward a subjectivity-aware mindset, where variation is recognised as meaningful signal rather than noise. Second, disagreement among annotators reflects the inherent complexity of the domain and exposes the limits of any single perspective. Third, students must recognise their own role in constructing datasets, acknowledging that data gathering is inherently subjective and carries biases that propagate into downstream model outcomes.

\begin{figure*}[t]
\centering
\resizebox{\textwidth}{!}{%
\begin{tikzpicture}[
    box/.style={rectangle, draw, rounded corners, minimum width=3.2cm, minimum height=2.8cm, align=center, font=\small},
    arrow/.style={->, thick, >=stealth},
    milestone/.style={circle, draw, fill=white, inner sep=2pt, font=\scriptsize},
    label/.style={font=\scriptsize, align=center}
]

\node[box, fill=gray!10] (before) at (0, 0) {
    \textbf{Before annotation}\\[2pt]
    \textit{Traditional mindset}\\[4pt]
    • Single ground truth\\
    • Disagreement = error\\
    • Labels are objective
};

\node[box, fill=gray!10] (after) at (10.5, 0) {
    \textbf{After annotation}\\[2pt]
    \textit{Subjectivity-aware}\\[4pt]
    • Multiple valid labels\\
    • Disagreement = signal\\
    • Labels are subjective
};

\draw[arrow] (before) -- (after);

\node[milestone] (m1) at (2.7, 0) {};
\node[milestone] (m2) at (4.2, 0) {};
\node[milestone] (m3) at (5.7, 0) {};
\node[milestone] (m4) at (7.5, 0) {};

\node[label, above=8pt of m1] {Annotation\\ambiguity};
\node[label, above=8pt of m2] {Data\\quality};
\node[label, above=8pt of m3] {Bias and\\fairness};
\node[label, above=8pt of m4] {Implementation\\barriers};

\node[label, below=12pt of $(m2)!0.5!(m3)$] {\textbf{Pedagogical effectiveness}};

\end{tikzpicture}
}
\caption{The learning persona trajectory. Students ideally move from a traditional mindset (left) to a subjectivity-aware mindset (right).}
\label{fig:learning-persona}
\end{figure*}
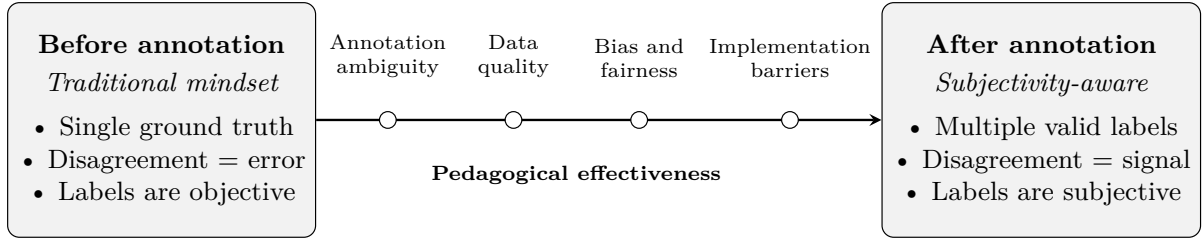

The framework directly maps to our two research questions. Importantly, the learning persona is a theoretical ideal and serves as an analytical guide rather than a grading rubric. Figure~\ref{fig:learning-persona} summarises the learning trajectory. Students should ideally develop or deepen a subjectivity-aware mindset, regardless of their prior assumptions about annotation.

The framework operationalises the three assumptions through five key concepts:
\begin{enumerate}
    \item Annotation ambiguity: The persona understands that his individual interpretation directly influences annotations (AA2) and realises that discussing disagreements with peers clarifies why different people assign different labels to the same image (AA3). The persona accepts that group consensus on the annotation results is not always a possibility. For a specific skin lesions, there may exist multiple valid annotations. The student shifts from a single truth to multiple valid ones.
    \item Data quality: The persona acknowledges the link between personal interpretation and data labeling (DQ1) and becomes aware of the challenges involved in creating training data (DQ2). The student questions the concept of ``Ground Truth''. The persona realises that datasets are constructed aggregations of human biases and not objective realities.
    \item Bias and fairness: The student connects specific visual features (such as hair) to possible misclassifications by the AI model (BF1 and BF2) and understands that analysing disagreements helps identify cues difficult for models to learn (BF3). The persona understands that AI bias often starts during the data collection phase (not solely because of the algorithm) and that a model trained on ``consensus'' labels may reinforce the majority's biases, overlooking minority perspectives, or ambiguous edge cases.
    \item Implementation barriers: The activity provides clear annotation instructions (IB1) and an accessible labelling tool (IB2). In this context, the persona feels confident in his ability to label skin lesions (IB4), although he occasionally experiences discomfort when viewing medical images (IB3). The student develops confidence in his own interpretive judgement, even in the absence of expert calibration.
    \item Pedagogical effectiveness: The student rates the activity as more effective than traditional lectures in understanding bias (PE2) and would recommend it to others (PE3). The student was also more motivated to understand AI concepts after finishing the task (PE4). The persona sees the link between the creation of data and the real-world impact of AI.
\end{enumerate}

 We use the persona as an analytical perspective against which to evaluate both quantitative survey data and qualitative feedback.  

\section{Results}\label{results}
    In total, 43 students completed the complete survey: 30 from Fontys University of Applied Sciences and 13 from IT University of Copenhagen.

\subsection{Substantial familiarity gains across key concepts}\label{quant_pre_post}
    \begin{table*}[htbp]
\centering
\caption{Pre- and post-task familiarity with key concepts. Scale: 0 = Never heard of it, 1 = Heard of it but did not understand it, 2 = Basic familiarity, 3 = Good understanding, 4 = Fully understood it. Values represent pre/post percentages. Bold post values indicate an increase over the corresponding pre value.}
\label{tab:familiarity_pre_post}
\begin{tabular}{p{0.30\textwidth} *{5}{>{\centering\arraybackslash}p{0.1\textwidth}}}
\toprule
Concept & 0 & 1 & 2 & 3 & 4 \\
\midrule
Subjectivity in data annotation & 9.5/0.0 & 16.7/0.0 & 23.8/23.8 & 31.0/\textbf{40.5} & 19.0/\textbf{35.7} \\
Inter-annotator agreement measures & 26.2/2.4 & 14.3/0.0 & 26.2/\textbf{28.6} & 26.2/\textbf{45.2} & 7.1/\textbf{23.8} \\
Bias in ML models & 11.9/0.0 & 14.3/7.1 & 28.6/14.3 & 33.3/\textbf{47.6} & 11.9/\textbf{31.0} \\
Data quality assessment & 7.1/0.0 & 19.0/2.4 & 35.7/16.7 & 23.8/\textbf{45.2} & 14.3/\textbf{35.7} \\
\bottomrule
\end{tabular}%
\end{table*}

Table \ref{tab:familiarity_pre_post} presents the familiarity of participants with our four core concepts prior to and after the annotation task.

\paragraph{Subjectivity in data annotation}
Before the annotation task, approximately one-quarter of the students reported little or no prior exposure to the concept of subjectivity in data annotation. After the intervention, no student remained at the lowest levels. The majority of the students became aware of it, and almost everyone moved to the highest scale levels.

\paragraph{Inter-annotator agreement measures}
A substantial group of students reported little or no prior familiarity with inter-annotator agreement measures. After completion of the task, most students report an improved knowledge of these measurement techniques.

\paragraph{Bias in ML models} 
Students report some awareness of bias in ML models before the annotation task has started. After the task, we see that no student remained completely unfamiliar with the concept, and those reporting good or full understanding increased from 45.2\% to 78.6\%. Nevertheless, approximately one-fifth still indicated only partial understanding.

\paragraph{Data quality assessment}
The students entered with limited background knowledge on data quality, yet after the intervention, nearly all reported a greater understanding.

\subsection{Quantitative findings on survey constructs}
    Students acknowledged that annotations are influenced by personal interpretation and appreciated opportunities to talk about differing views. Nonetheless, most described their own group’s annotations as largely aligned, so they did not encounter the kind of disagreement that is essential to the learning objective.

Most students recognised that individual judgement influences how data are labelled and that producing training datasets is difficult. They were able to link differences in annotations to unreliable model behaviour and noticed that visual attributes such as hair can lead to incorrect classifications. The activity provided them with a tangible understanding of why AI systems can generate untrustworthy predictions.

Most students considered the instructions and tools sufficient and felt self-assured when assigning labels, although a small subset reported discomfort when looking at the skin lesion images. They rated the annotation activity as more effective than standard lectures for learning about bias, and the majority supported integrating similar exercises into other courses. Motivation and intrinsic interest also rose, but roughly one-quarter of students remained neutral, indicating that the activity does not appeal to everyone equally.

Table \ref{tab:likert_combined} presents the Likert-scale response distributions in all five constructs evaluated in our survey. These constructs were designed to measure student perceptions about annotation ambiguity, data quality, bias and fairness, implementation barriers, and pedagogical effectiveness. In the following subsections, we analyse the quantitative findings for each construct.

\subsubsection{Annotation ambiguity}\label{AA}
    We found that while most students reported consensus within their groups, a substantial minority noted limited consistency among annotators (AA1). The majority indicated that discussing these disagreements helped them understand why different individuals assign varying labels to the same image (AA3). Furthermore, most of the students acknowledged that personal interpretation plays a role in annotating skin lesions (AA2). In their open-ended responses, several students reflected on the thoroughness of their own annotation process, requesting calibration methods and clearer guidelines to improve consistency (See AF1 and IB5 in Fig. \ref{fig:heatmap}). In other words, students treated inconsistency as a solvable problem rather than recognising it as an inherent feature of the task. This contrasts with our learning persona, who accepts that consensus is not always achievable and treats multiple valid annotations as normal. Although most of the students reported group consensus (AA1), this does not imply that they considered consensus universally possible or desirable. Qualitative findings (Section \ref{trad_modern_view}) show mixed opinions: some treated disagreement as a problem to fix, others as an inherent feature of the task.
    
\subsubsection{Data quality}
    We observed that most of the students recognised the link between their personal interpretations and their data annotations (DQ1). Additionally, the task made them realise that creating training data can be challenging (DQ2).

\subsubsection{Bias and fairness}
    We found that most of the students understood why AI models produce uncertain predictions after comparing their annotations with those of their peers (BF1). A strong majority recognised that visual features such as hair contribute to classification errors (BF2), and many agreed that analysing disagreements helped them identify visual cues that are difficult for models to learn (BF3). In general, the annotation task provided meaningful insight into why AI systems can yield unreliable classifications (BF4).

\subsubsection{Implementation barriers}
    We observed that the activity provided clear instructions (IB1) and an easy-to-use labelling tool (IB2), both of which most students rated positively. We found that most students felt confident in their ability to label skin lesions (IB4), and while some reported mild discomfort when viewing medical images (IB3), this did not appear to hinder their engagement with the task. This broadly aligns with our learning persona, who likewise finds the instructions clear, the tool intuitive, and feels confident in labelling. We also found that several students suggested improvements such as providing example annotations or calibration exercises (IB5). These suggestions show that students identified practical challenges in the annotation process, but attributed them to a lack of guidance or tooling rather than to inherent subjectivity of the task.

\subsubsection{Pedagogical effectiveness}
    We found that the majority of the students were motivated by an interest in data annotation that extends beyond the basic course requirements (PE1). This positive sentiment carried over to comparisons with traditional lectures, and most of the participants agreed that the labelling activity was more effective in helping them understand bias in AI (PE2). Consistent with this, most participants suggested embedding a data annotation assignment within other AI or data science courses to help students grasp the roles of subjectivity and bias (PE3). In addition, the labeling activity increased the motivation of most participants to understand the concepts of AI (PE4).

\begin{table*}[htbp]
\centering
\caption{Likert-scale responses for all constructs. SD = Strongly Disagree (1), D = Disagree (2), MD = Mildly Disagree (3), N = Neutral (4), MA = Mildly Agree (5), A = Agree (6), SA = Strongly Agree (7). Values are percentages.}
\label{tab:likert_combined}
\renewcommand{\arraystretch}{1.15}
\scriptsize
\begin{tabular}{p{0.40\textwidth} *{7}{>{\centering\arraybackslash}p{0.05\textwidth}}}
\toprule
Question & SD & D & MD & N & MA & A & SA \\
\midrule
\multicolumn{8}{l}{\textit{Annotation Ambiguity (AA)}} \\
1. The annotations produced by my group were consistent with each other. & 7.1 &  & 14.3 & 9.5 & 23.8 & \textbf{40.5} & 4.8 \\
2. My individual interpretation played a role in how I labelled the skin lesions. & 2.4 &  &  & 9.5 & 23.8 & \textbf{38.1} & 26.2 \\
3. Discussing annotation disagreements with my peers helped me understand why different people can assign different labels to the same image. &  & 4.8 &  & 21.4 & 23.8 & 21.4 & \textbf{28.6} \\
\addlinespace[4pt]
\multicolumn{8}{l}{\textit{Data Quality (DQ)}} \\
1. The annotation task helped me understand how personal interpretation affects data labeling. &  &  &  & 4.8 & 19.0 & \textbf{50.0} & 26.2 \\
2. Doing the annotation task made me more aware of the challenges in creating training data. &  &  & 2.4 & 2.4 & 14.3 & \textbf{47.6} & 33.3 \\
\addlinespace[4pt]
\multicolumn{8}{l}{\textit{Bias and Fairness (BF)}} \\
1. Reviewing disagreements between my labels and my peers' labels helped me understand why AI models might produce uncertain or conflicting predictions. &  & 2.4 & 2.4 & 14.3 & \textbf{28.6} & 26.2 & 26.2 \\
2. The presence of hair in training images contributes to misclassifications by the AI model. &  &  & 2.4 & 9.5 & 21.4 & \textbf{35.7} & 31.0 \\
3. Reflecting on why I disagreed with others helped me identify visual cues in skin lesions that may be difficult for an AI model to learn. & 2.4 & 4.8 & 4.8 & 16.7 & \textbf{26.2} & \textbf{26.2} & 19.0 \\
4. The annotation task gave me insight into why AI models can produce unreliable classifications. &  &  &  & 4.8 & 19.0 & \textbf{40.5} & 35.7 \\
\addlinespace[4pt]
\multicolumn{8}{l}{\textit{Implementation Barriers (IB)}} \\
1. The instructions for the annotation task were clear and easy to follow. &  &  & 19.0 & 11.9 & 19.0 & 21.4 & \textbf{28.6} \\
2. The annotation tool was easy to use and did not interfere with my ability to complete the task. &  & 4.8 & 11.9 & 19.0 & 19.0 & \textbf{26.2} & 19.0 \\
3. Viewing the skin lesion images caused me emotional discomfort or distress. & 14.3 & \textbf{21.4} & 7.1 & 9.5 & 7.1 & 19.0 & \textbf{21.4} \\
4. I felt confident in my ability to accurately label the skin lesion images. &  &  & 21.4 & 11.9 & \textbf{31.0} & 19.0 & 16.7 \\
\addlinespace[4pt]
\multicolumn{8}{l}{\textit{Pedagogical Effectiveness (PE)}} \\
1. I was interested in learning about data annotation beyond just completing the course requirement. &  & 2.4 & 11.9 & 21.4 & \textbf{31.0} & 19.0 & 14.3 \\
2. Compared to traditional lectures, the labeling activity was more effective in helping me understand bias in AI. &  & 2.4 & 9.5 & 19.0 & \textbf{28.6} & 16.7 & 23.8 \\
3. I would recommend incorporating a data annotation task into other AI or data science courses to help students understand subjectivity and bias. & 2.4 & 4.8 & 7.1 & 11.9 & \textbf{31.0} & 19.0 & 23.8 \\
4. The labeling activity increased my motivation to understand AI concepts. & 2.4 & 2.4 & 11.9 & 19.0 & 23.8 & \textbf{28.6} & 11.9 \\
\bottomrule
\end{tabular}%
\end{table*}
    
\subsection{Quantitative finding on inter-annotator agreement metrics}
    Although the Fontys students computed Fleiss's $\kappa$  themselves and the results indicated disagreement within the group, they did not reflect this in their survey responses, where they reported having reached consensus.

\begin{figure}[!ht]
    \centering
    \includegraphics[width=0.4\textwidth]{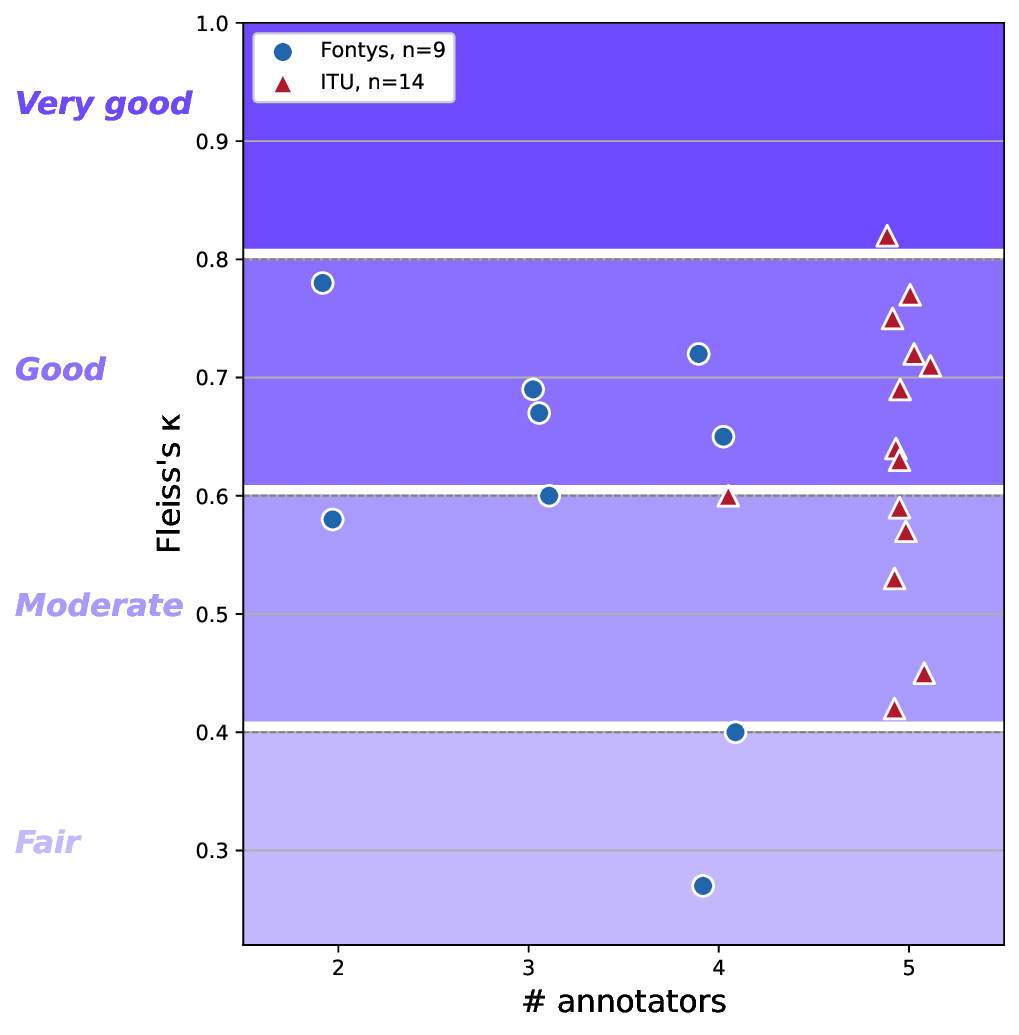}
    \caption{Inter-annotator agreement (Fleiss's $\kappa$) by group size for Fontys and Copenhagen ITU. Shaded bands represent $\kappa$ interpretation categories.}
    \label{fig:kappa_itu_fontys}
\end{figure}

For each annotation group at both institutions, Fleiss's $\kappa$ was computed (See Fig. \ref{fig:kappa_itu_fontys}, Appendix \ref{kappa_tables}). In Fontys, the nine groups yielded a mean $\kappa$ of 0.60, ranging from 0.27 (Fair) to 0.78 (Good). At ITU, the fourteen groups produced a mean $\kappa$ of 0.64, ranging from 0.42 (Moderate) to 0.82 (Very good). In general, the combined mean $\kappa$ in all 23 groups was 0.62, indicating moderate agreement. The moderate level of agreement indicates that the intended degree of subjectivity is indeed present in the dataset.
To contextualise these metrics, we compared them with students' self-reported consensus. Although most of the students indicated that their group had reached consensus, the values of $\kappa$ suggest that this agreement was only moderate in practice. At Fontys, two groups scored in the Fair range ($\kappa < 0.40$), but neither group appears to have reported notable disagreement in the survey. This gap between perceived and measured agreement is noteworthy: it suggests that students may define consensus more loosely than the metric does.
This result relates to the wider issue of what is addressed within the AA construct (Section \ref{AA}). Students who reported group consensus despite moderate agreement may treat any convergence as success, failing to recognise that the remaining disagreement reflects the task's subjectivity. Even when the Fontys students calculated $\kappa$ themselves, the issue remained, showing that computing the metric alone does not change their conceptual framework.

\subsection{Qualitative findings from open-ended items}
    In our thematic analysis of the open-ended responses, we found that students recognised the significant role of personal interpretation in annotation, but rather than engaging with it, they attempted to eliminate it. They found the activity itself repetitive, but enjoyed discussing results and sharing ideas afterwards. What kept them engaged was not the activity itself, but seeing how it connects to real-world. Overall, a consistent pattern emerged: students recognized the ambiguity, explicitly pointed it out, and then requested that it be removed. In the subsections that follow, we examine these findings in more depth.

\subsubsection{Most significant insights during annotation }
    We found that subjectiveness was the most frequent theme ($\bar{n}=22.5$), as the students saw how personal interpretation shapes labelling. We also noted mentions of calibration methods ($\bar{n}=13$) and agreement issues ($\bar{n}=12.5$). This shows that the students recognised the challenges of keeping annotations consistent between groups. A smaller group linked their insights to AI models ($\bar{n}=3.5$). They saw how annotation subjectivity affects model training. The students identified subjectivity as the main theme, and their reaction to it is to seek calibration and agreement.

\subsubsection{Suggestions for improving the task}
    The most frequent suggestion was to make annotations more consistent, for example through clearer guidelines, calibration exercises, or reference images ($\bar{n}=20$). This response reveals a central pedagogical tension: students who had just experienced annotation subjectivity firsthand sought to eliminate it rather than learn from it. One student noted, clearer criteria could help make annotations ``more consistent'' , yet also acknowledged, ``I understand also that this was probably the learning journey.'' This remark suggests that at least some students grasped the paradox, even if most did not. A second cluster of responses concerned practical implementation: students requested simpler tooling ($\bar{n}=10$), such as pre-configured notebooks or alternatives to Label Studio, and additional teaching support ($\bar{n}=7$), such as more coding walkthroughs. While these suggestions are actionable, they do not address the conceptual core of the exercise. A smaller subset of students explicitly called for greater transparency about the role of subjectivity in the task ($\bar{n}=3.5$), and a few reported that the task required no improvement ($\bar{n}=2$).

\subsubsection{Least enjoyable part of the task}
    We identified personal discomfort as the dominant complaint ($\bar{n}=23.5$). This occurred when viewing images of skin lesions. We found that this confirmed the quantitative findings on emotional distress (IB3). We saw that the second drawback was repetitiveness ($\bar{n}=12$). Students found labeling large image sets monotonous. We noted that smaller clusters refer to the annotation tool ($\bar{n}=4.5$), the task being time consuming ($\bar{n}=2.5$), and problems with group work ($\bar{n}=2$).

\subsubsection{Most enjoyable part of the task}
    We found that analysing the results was most appreciated ($\bar{n}=21$). We noted that the second enjoyable element was the group exchange ($\bar{n}=14.5$). We also saw that learning progress was valued ($\bar{n}=8.5$).

\subsubsection{What students found interesting about skin lesions}
    We found that the most frequent theme was usefulness in real life ($\bar{n}=11.5$). We also saw that the data ($\bar{n}=7$) and the annotation task itself ($\bar{n}=6$) were considered interesting. ML ($\bar{n}=5.5$) and trustworthy AI ($\bar{n}=5$) received similar interest.

\subsection{Did students take on the traditional or modern view of subjectivity of human annotations?}\label{trad_modern_view}

\begin{figure*}[!ht]
    \centering
    \includegraphics[width=0.9\textwidth]{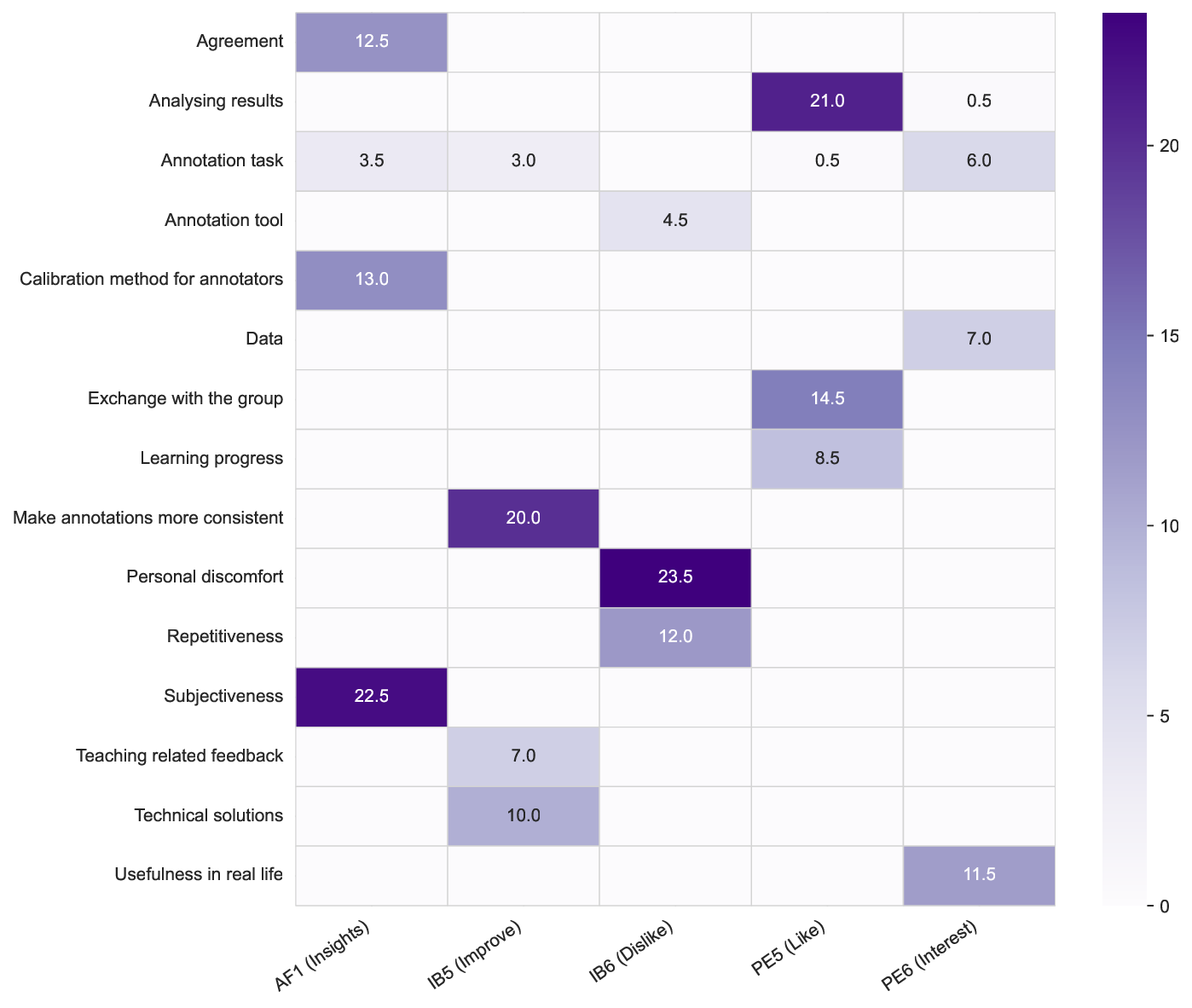}
    \caption{Heatmap of average annotation label frequencies ($\bar{n}$ = average across two annotators) across coded themes ((rows, ordered alphabetically)) and open-ended survey questions (columns: AF1, IB5, IB6, PE5, PE6). Darker cells indicate higher average. The three most frequent themes per survey question are shown.}
    \label{fig:heatmap}
\end{figure*}

To understand the conceptual takeaways of the students in light of recent research on subjective annotation, we analysed their responses to implicitly adopted views on subjectivity aggregated over all open-ended questions as presented in Fig.~\ref{fig:heatmap}. Specifically, we examined whether students recognised the subjective nature of data annotation or instead focused on agreement. Of the responses, $\bar{n}=22.5$ referred to the subjectivity of annotations, while $\bar{n}=12.5$ focused on agreement. This suggests that a substantial proportion of students reflected on the role of subjectivity in the annotation process rather than viewing the annotation solely through the lens of consensus.

We further analysed how students framed annotator agreement. The responses were categorised according to whether they reflected the traditional agreement paradigm, which emphasises maximising agreement and treats disagreement primarily as annotation noise, or the more recent disagreement paradigm, which recognises that differences between annotators can arise from legitimate variation in interpretation and therefore carry valuable information~(cf. Section~\ref{Relatedwork}). Among the responses discussing agreement, $\bar{n}=8.5$ explicitly characterised agreement as desirable (``agreement good''), whereas only $\bar{n}=1.5$ framed disagreement negatively (``agreement bad''). In contrast, among the responses discussing subjectivity, $\bar{n}=2$ explicitly described subjectivity as beneficial (``subjectiveness good''), while $\bar{n}=4.5$ viewed it negatively (``subjectiveness bad''). The remaining responses acknowledged subjectivity or agreement without expressing a clear stance. Overall, these findings indicate that while many students recognised the subjective nature of annotation, only a smaller subset explicitly adopted the perspective that subjectivity and disagreement between annotators can be valuable rather than merely sources of error.


\section{Discussion and conclusions}\label{Discussion}
    Across the two institutions, our results show a productive pedagogical tension. On the one hand, students who annotated skin lesion images reported substantial gains in their understanding of subjectivity, bias, and data quality (RQ1). On the other hand, when asked how the task could be improved, the most frequent suggestion was to reduce the very disagreement that drove this learning (RQ2). This paradox is, we argue, not a failure of the intervention but it is an actionable pedagogical finding. It shows that students have begun to notice the subjectivity of data but have not yet internalised that disagreement is a signal of domain complexity rather than a defect to be corrected.

\subsection{Successes and challenges}
    


    \subsubsection{Making subjectivity visible through the agreement paradigm}
    The agreement paradigm (as discussed in Section~\ref{sec:agreement-disagreement-paradigm}) makes subjectivity of data annotation visible to students by treating annotation quality as the degree to which annotators converge on a single ground truth. Measuring inter-annotator agreement using Fleiss $\kappa$, our task deliberately asked students to quantify their agreement. In this sense, the annotation activity first taught the students the traditional perspective of data annotation quality. Quantitatively, we observed an increased self-reported familiarity with inter-annotator agreement measures. The students highly valued the analysis of the of their annotation agreement and the respective group exchange (see PE5 in Figure~\ref{fig:heatmap}). Moreover, students reported greater knowledge of subjectivity and agreement between annotators (Table \ref{tab:familiarity_pre_post}). Greatest evidence of our task's success to make subjectivity in data annotation visible to students was, that subjectiveness was the among the most commonly reported topics in across all survey responses (see AF1 in Figure~\ref{fig:heatmap}, n=22.5). Critically, we observe that students, when confronted with their different assessments, reported on group consensus when there was only moderate measurable agreement (AA1) as per their group's Fleiss' $\kappa$. We attribute this skewed perception of consensus to the agreement paradigm that frames any disagreement as undesirable. This mirrors our observation in Section~\ref{sec:data-annotation-in-teaching} that agreement-based exercises in education implicitly present consensus as the endpoint not the start of a discussion. Especially for teaching bias and fairness and making students aware of responsible AI, the agreement paradigm is not sufficient. A model trained only on aggregated ground truth labels encodes the majority's interpretation and erases minority perspectives. Students reach this insight only when they shift from the agreement paradigm to the disagreement paradigm.

    \subsubsection{Making subjectivity useful through the disagreement paradigm}
    The central learning goal of the annotation task was to increase awareness of students towards the value of subjectivity, i.e. to move their understanding of subjectivity in data annotation from the agreement paradigm to the disagreement paradigm as illustrated in Figure~\ref{fig:learning-persona}. Our findings show that a few students have fully embraced this paradigm shift, describing subjectivity as beneficial because it carries signal, while others portrayed subjectivity more negative as noise. In between, there is evidence of a partial paradigm shift among many students. They understand there is not one single right answer but that differences in annotations are often justified individually. Evidence of this is found in the fact that most students acknowledged that personal interpretation shapes their labels (AA2), and they found that discussing disagreements clarified why people label the same image differently (AA3). 
    Moreover, students recognised that the subjectivity of annotations affects the accuracy of AI model outputs. This insight is concrete evidence that our manual data annotation task sensitizes students to topics of trustworthy AI. They are aware of subjectivity in data labels and therefore less likely to adopt a false sense of objectivity in data and AI model outputs. Students arrived at these ideas without being introduced to the literature on the disagreement paradigm, which suggests that hands-on annotation can surface the intuitions behind it on its own at least to some degree. On the other hand, negative framing of disagreement included suggestions to reduce it through clearer guidelines, calibration rounds, or example annotations (IB5). In terms of the taxonomy by Sandri et al.~\cite{sandri:2023}, students tended to attribute disagreement to noise-driven sources such as unclear instructions. They rarely considered that part of it was genuinely subjective. Our annotation task deliberately involves both kinds of disagreement. Learning to tell them apart is one of the competences that future iterations of the course should actively investigate.

    The calls for consistency (IB5) pose a pedagogical challenge for us. Providing the requested calibration would ironically undermine the very goal of the exercise: allowing students to experience annotation ambiguity directly. However, if designed carefully, such a change could emphasize the difference between misunderstandings and genuine subjectivity which both without further inspection, appear as annotator disagreement. It would be important to keep the struggle with inconsistency as the lesson to learn for students. Interestingly some students demonstrated awareness of these different causes for disagreement, noting that ambiguity was likely the core of the learning journey. This insight validates the teaching method but highlights that many students have not yet fully internalised the principle that subjectivity can carry important signal. Consequently, teachers may need to explicitly emphasize the value of disagreement, ensuring students do not view the lack of consensus as a failure of the task or their skills.

    \subsubsection{Student discomfort} A substantial proportion of the students noted feeling uneasy about the medical imagery ($\bar{n}=23.5$), see IB6 in Fig. \ref{fig:heatmap}. This approach can enhance educational value, but only if it is framed appropriately (see Section \ref{guidelines}). Otherwise, the experience risks overshadowing the learning.
    
    \subsubsection{Technical friction} Students frequently noted technical and procedural barriers. The students reported that the annotation tool setup was cumbersome and the manual labelling process was ``boring'' and ``slow''. These frustrations often overshadowed the learning potential for some participants, who suggested simpler tools (e.g., ``a python script'') or pre-built notebooks to focus purely on the decision-making logic.

    \subsubsection{Findings through a didactic perspective}
    Our findings provide evidence that students experience learning by doing as more effective than traditional lectures alone. Raschka's paper supports this observation: the author states that absorbing knowledge from lectures is ``necessary but not sufficient''.  Project work on a realistic topic ``can motivate students to engage in topics more deeply'' \cite{Raschka2021-qi}. The results of our study suggest a comparable increase in motivation and engagement. Students reported an interest in data annotation that went beyond merely fulfilling the course requirements (PE1). Moreover, the labelling activity increased their motivation to understand AI concepts (PE4).

    We have seen that analysing annotations and group exchange were most valued by our students. Pease \& Kuhn \cite{Pease2011-rd} found that social collaboration itself did not account for the effectiveness of PBL. The authors explain this effect by the goal-oriented activity framework it offers. Our annotation task incorporates precisely this structure, beginning with the initial question (how many hairs are visible?), progressing through intermediate steps, and ending with the conclusion (the impact of annotations on model behaviour). Our survey data show that most students agreed that reviewing disagreements between labels helped them understand why AI models might produce uncertain or conflicting predictions (BF1), and that reflecting on why they disagreed with peers helped them identify visual cues that are difficult for an AI model to learn (BF3). These findings aligns with Pease \& Kuhn's explanation: what students valued most was not the collaboration per se, but the evidence-based reasoning it enabled.

\subsection{Pedagogical takeaways}\label{guidelines}

    \subsubsection{Ensuring sufficient interpretive ambiguity} The annotation task should have sufficient interpretive ambiguity so that disagreement naturally arises among students within a group (labeling cats and dogs is a counter-example). This confronts students with conflicting human judgments, transforming the labeling process from a mechanical exercise into a pedagogical means. The developer of the annotation task has to ensure that the label score is in a grey area. However, it must be ensured that students do not mistake misunderstandings resulting from missing annotation guidelines or task descriptions for subjectivity. 

    \subsubsection{Balancing repetitiveness and reflection} We observed a notable tension in the results between repetitiveness and analysis of results. Repetitiveness appeared as a key drawback (IB6, $\bar{n}=12$). However, analysing the results was the most pleasant element (PE5, $\bar{n}=21$). This suggests that students value reflection rather than mechanical labelling. This finding is of importance for the design of annotation tasks. We propose optimising the workflow to ease annotation. Reducing the total image count could also help. Consider lowering the minimum requirement from 100 to 50 images. Flexibility remains essential. Students who wish to annotate more should be allowed to do so.

     \subsubsection{Mitigating discomfort} The learning persona trajectory assumes that students will encounter only occasional discomfort and normalise it quickly. Feedback indicates that unease with medical imagery was the most frequent concern. Educators should anticipate varying reactions when working with medical data. We suggest considering medical framing before use. For example, explain the relevance of images to melanoma detection. Converting images to greyscale before annotation may also help reduce visual intensity \cite{Karunakaran2019-mv}. Structured preparation and brief debriefing sessions could also be considered to help students process these experiences.

     \subsubsection{Closing the loop between annotations and model output} We showed that students recognise the link between annotation data and model performance. At ITU, students already implement this loop: they train a simple model on their aggregated annotation data and subsequently analyse the predictions. We recommend adopting this approach more systematically. We invite students to answer specific questions. For instance, ``Where did the model fail?'' ``Was it due to ambiguous data or annotator disagreement?'' This approach requires that annotations influence the model output. We demonstrated this in our earlier paper, in which we showed that annotations from various sources affected performance \cite{raumannsr-melba}.

\subsection{Limitations}
    The study has some limitations. First, the modest sample size (N=43) limits generalisability. Second, pre-task familiarity was measured retrospectively after activity completion, relying on students' reconstructed memory rather than baseline assessment. Third, while the task design remained consistent, the annotation tools varied across institutions: Fontys students used Label Studio or custom Jupyter notebooks, whereas ITU students were free to choose their preferred tool, with many opting for Excel. This tool heterogeneity may affect reproducibility in other institutional contexts. Finally, our study spans two institutions, with different pedagogical designs. These contextual differences, as well as variations in group composition, were not systematically analysed and may influence the results.
    
\subsection{Future work}
    An important direction for future research is to explore how students might rely on large language models (LLMs) to carry out annotation tasks, rather than performing these tasks by hand. This raises two key questions: First, does relying on LLM-generated annotations hurt students' learning? Manual annotation helps students develop close reading and critical thinking skills, so bypassing it could reduce the educational value of the task. Second, how can course design respond? Future versions of the course could encourage genuine engagement through task design or reflection exercises. At the same time, LLM-based annotation is an emerging research field in itself, so it could be introduced as an explicit learning goal rather than treated only as a problem to prevent. Examining these questions will help keep annotation-based teaching effective as automated tools continue to become more widely available.

    Another valuable direction is investigating the gap between students' perceived agreement and actual statistical agreement. Our findings suggest a ``consensus illusion'': students reported agreement despite only moderate Fleiss's $\kappa$ values. Follow-up studies could, for example, use recorded group discussions to determine whether this gap originates from metric confusion, social dynamics, or beliefs about ground truth. Understanding this ``consensus illusion'' would help educators design annotation activities that frame disagreement as informative rather than erroneous.

\newpage
\acks{This work was supported by the Netherlands Organisation for Scientific Research, grant no. 023.014.010. VC is supported by Novo Nordisk, grant number NNF24OC0092612. We sincerely thank the students from Fontys University of Applied Sciences Venlo and the IT University of Copenhagen for their participation. Their engagement in annotating skin lesion images provided essential data for our research. We are particularly grateful to those who shared their insights through the post-task survey. We also acknowledge the teaching staff at both institutions for integrating the annotation task into the regular curriculum. Special thanks go to our dogs Max Moritz and Beau Jeanne d'Arc, and cats Pixel and Dot, for graciously agreeing to pose for the camera.}

\ethics{The work follows appropriate ethical standards in conducting research and writing the manuscript, following all applicable laws and regulations regarding treatment of animals or human subjects.}

\coi{The authors declare that they have no known competing financial interests or personal relationships that could have influenced the work reported in this paper. The software and survey server for conducting the online survey were provided free of charge by the SoSci Survey GmbH. The SoSci Survey GmbH had no influence on the content of the study.}

\data{The data and code are publicly available through \url{https://github.com/raumannsr/DAPH}}

\newpage
\bibliographystyle{plain}
\bibliography{bibliography/refs_ralf,bibliography/refs_gerard,bibliography/refs_veronika,bibliography/refs_theresa}

\newpage
\begin{appendix}
    \section{Post-test DAPH}
    \subsubsection{Annotation ambiguity (AA)}
    \begin{enumerate}[label=AA\arabic*, leftmargin=*]
        \item \hfill RQ1 \\
        \textbf{The annotations produced by my group were consistent with each other.}\\
        Strongly Disagree \textrightarrow{} Strongly Agree
        \item \hfill RQ1 \\
        \textbf{My individual interpretation played a role in how I labelled the skin lesions during the assignment.} \\
        Strongly Disagree \textrightarrow{} Strongly Agree
        \item \hfill RQ1 \\
        \textbf{Discussing annotation disagreements with my peers helped me understand why different people can assign different labels to the same image.} \\
        Strongly Disagree \textrightarrow{} Strongly Agree
    \end{enumerate}

\subsubsection{Data quality (DQ)}
    \begin{enumerate}[label=DQ\arabic*, leftmargin=*]
        \item \hfill RQ1 \\
        \textbf{The annotation task helped me understand how personal interpretation affects data labeling.} \\
        Strongly Disagree \textrightarrow{} Strongly Agree
        \item \hfill RQ1 \\
        \textbf{Doing the annotation task made me more aware of the challenges in creating training data.} \\
        Strongly Disagree \textrightarrow{} Strongly Agree
    \end{enumerate}

\subsubsection{Bias and fairness (BF)}
    \begin{enumerate}[label=BF\arabic*, leftmargin=*]
        \item \hfill RQ1 \\
        \textbf{    Reviewing disagreements between my labels and my peers' labels helped me understand why AI models might produce uncertain or conflicting predictions.}\\
        Strongly Disagree \textrightarrow{} Strongly Agree
        \item \hfill RQ1\\
        \textbf{The presence of hair in training images contributes to misclassifications by the AI model.} \\
        Strongly Disagree \textrightarrow{} Strongly Agree
        \item \hfill RQ1 \\
        \textbf{Reflecting on why I disagreed with others helped me identify visual cues in skin lesions that may be difficult for an AI model to learn.} \\
        Strongly Disagree \textrightarrow{} Strongly Agree
        \item \hfill RQ1 \\
        \textbf{The annotation task gave me insight into why AI models can produce unreliable classifications.} \\
        Strongly Disagree \textrightarrow{} Strongly Agree
    \end{enumerate}

\subsubsection{Covering AA, DQ and BF}
    \begin{enumerate}[label=AF\arabic*, leftmargin=*]
        \item \hfill RQ1 \\
        \textbf{What were the most significant insights or findings your group discovered during the annotation process?} \\
        Free text.
    \end{enumerate}

\subsubsection{Implementation barriers (IB)}
    \begin{enumerate}[label=IB\arabic*, leftmargin=*]
        \item \hfill RQ2 \\
        \textbf{The instructions for the annotation task were clear and easy to follow.} \\
        Strongly Disagree \textrightarrow{} Strongly Agree
        \item \hfill RQ2 \\
        \textbf{The annotation tool was easy to use and did not interfere with my ability to complete the task.} \\
        Strongly Disagree \textrightarrow{} Strongly Agree
        \item \hfill RQ2 \\
        \textbf{Viewing the skin lesion images caused me emotional discomfort or distress.}\\
        Strongly Disagree \textrightarrow{} Strongly Agree
        \item \hfill RQ2 \\
        \textbf{I felt confident in my ability to accurately label the skin lesion images.} \\
        Strongly Disagree \textrightarrow{} Strongly Agree
        \item \hfill RQ2 \\
        \textbf{How would you improve the task? } \\
        Free text.
        \item \hfill RQ2 \\
        \textbf{What is the least enjoyable part of the task? } \\
    Free text.

    \end{enumerate}

\subsubsection{Pedagogical effectiveness (PE)}
    \begin{enumerate}[label=PE\arabic*, leftmargin=*]
        \item \hfill RQ2 \\
        \textbf{I was interested in learning about data annotation beyond just completing the course requirement.} \\
        Strongly Disagree \textrightarrow{} Strongly Agree
        \item \hfill RQ2 \\
        \textbf{Compared to traditional lectures, the labeling activity was more effective in helping me understand bias in AI.} \\
        Strongly Disagree \textrightarrow{} Strongly Agree
        \item \hfill RQ2 \\
        \textbf{I would recommend incorporating a data annotation task into other AI or data science courses to help students understand subjectivity and bias.} \\
        Strongly Disagree \textrightarrow{} Strongly Agree
        \item \hfill RQ2 \\
        \textbf{The labeling activity increased my motivation to understand AI concepts.} \\
        Strongly Disagree \textrightarrow{} Strongly Agree
        \item \hfill RQ2 \\
        \textbf{What is the most enjoyable part of the task?} \\
        Free text.
        \item \hfill RQ2 \\
        \textbf{What, if anything, did you find interesting about the topic of skin lesions?} \\
        Free text.
    \end{enumerate}

\section{Inter-annotator agreement Fontys and ITU}\label{kappa_tables}

\begin{table}[h!]
    \centering
    \caption{Available Fleiss's $\kappa$ results reported by Venlo students ($n = groupsize$), with $\kappa$: $< 0.20$ = Poor, $0.21-0.40$ = Fair, $0.41-0.60$ = Moderate, $0.61-0.80$ = Good and $0.81-1.00$ = Very good. \\}
    \label{tab:fleiss_results_venlo}
    \begin{tabular}{l c c c l}
        \toprule
        \textbf{Group} & \textbf{$n$} & \textbf{Fleiss' $\kappa$} & \textbf{Interpretation} \\
        \midrule
        1 & 2 & 0.58 & Moderate \\
        2 & 3 & 0.60 & Moderate \\
        3 & 3 & 0.67 & Good \\
        4 & 3 & 0.69 & Good \\
        5 & 2 & 0.78 & Good \\
        6 & 4 & 0.27 & Fair \\
        7 & 4 & 0.72 & Good \\
        8 & 4 & 0.40 & Fair \\
        9 & 4 & 0.65 & Good \\
        \bottomrule
    \end{tabular}
\end{table}
\begin{table}[!ht]
    \centering
    \caption{Fleiss's $\kappa$ results for Copenhagen student groups, computed from their annotations.}
    \label{tab:fleiss_results_copenhagen}
        \begin{tabular}{l c c c l}
        \toprule
        \textbf{Group} & \textbf{$n$} & \textbf{Fleiss' $\kappa$} & \textbf{Interpretation} \\
        \midrule
        1 & 4 & 0.60 & Moderate \\
        2 & 5 & 0.82 & Very good \\
        3 & 5 & 0.71 & Good \\
        4 & 5 & 0.45 & Moderate \\
        5 & 5 & 0.64 & Good\\
        6 & 5 & 0.42 & Moderate \\
        7 & 5 & 0.53 & Moderate \\
        8 & 5 & 0.69 & Good \\
        9 & 5 & 0.77 & Good \\
        10 & 5 & 0.57 & Moderate \\
        11 & 5 & 0.59 & Moderate \\
        12 & 5 & 0.72 & Good \\
        13 & 5 & 0.75 & Good \\
        14 & 5 & 0.63 & Good \\
        \bottomrule
    \end{tabular}
\end{table}

\newpage
\end{appendix}

\end{document}